\documentclass[referee,twocolumn]{aastex7}
\usepackage{graphicx}
\usepackage{indentfirst}
\usepackage{subfigure}
\usepackage{natbib}

\RequirePackage{xspace}
\RequirePackage{amsmath}
\RequirePackage{amsfonts}
\RequirePackage{amssymb}

\def\ifundefined#1{\expandafter\ifx\csname#1\endcsname\relax}

\makeatletter
\newcommand*{\rom}[1]{\expandafter\@slowromancap\romannumeral #1@}
\makeatother

\def\la{\mathrel{\hbox{\rlap{\hbox{\lower4pt\hbox{$\sim$}}}\hbox{$<$}}}}
\def\ga{\mathrel{\hbox{\rlap{\hbox{\lower4pt\hbox{$\sim$}}}\hbox{$>$}}}}

\newcommand{\be}{\begin{equation}}
\newcommand{\ee}{\end{equation}}

\newcommand{\bea}{\begin{eqnarray}}
\newcommand{\eea}{\end{eqnarray}}

\ifundefined{ensuremath}\def\ensuremath#1{\relax\ifmmode{#1}}
\else${#1}$\fi\else\relax\fi
\ifundefined{nuc}\def\nuc#1#2{\relax\ifmmode{}^{#1}{\protect\text{#2}}
\else${}^{#1}$#2\fi}\else\relax\fi

\ifundefined{ion}
\newcommand\ion[2]{#1$\;${%
\ifx\@currsize\normalsize\small \else
\ifx\@currsize\small\footnotesize \else
\ifx\@currsize\footnotesize\scriptsize \else
\ifx\@currsize\scriptsize\tiny \else
\ifx\@currsize\large\normalsize \else
\ifx\@currsize\Large\large
\fi\fi\fi\fi\fi\fi
\rmfamily\rom{#2}}\relax}%
\else\relax\fi

\newcommand{\kmps}{\ensuremath{\text{km}~\text{s}^{-1}}\xspace}

\newcommand{\msol}{\ensuremath{{\text{M}_\odot}}\xspace}

\newcommand{\phx}{\texttt{PHOENIX}\xspace}
\newcommand\phxO{\texttt{PHOENIX/1D}\xspace}

\makeatletter

\newcommand\Autoref[1]{\@first@ref#1,@}
\def\@throw@dot#1.#2@{#1}
\def\@set@refname#1{
    \edef\@tmp{\getrefbykeydefault{#1}{anchor}{}}%
    \def\@refname{\@nameuse{\expandafter\@throw@dot\@tmp.@autorefname}s}%
}
\def\@first@ref#1,#2{%
  \ifx#2@\autoref{#1}\let\@nextref\@gobble
  \else%
    \@set@refname{#1}
    \@refname~\ref{#1}
    \let\@nextref\@next@ref
  \fi%
  \@nextref#2%
}
\def\@next@ref#1,#2{%
   \ifx#2@ and~\ref{#1}\let\@nextref\@gobble
   \else, \ref{#1}
   \fi%
   \@nextref#2%
}

\makeatother
\usepackage{import}
\usepackage[colorinlistoftodos]{todonotes}

\usepackage[capitalize]{cleveref}

\xspaceaddexceptions{]}

\newcommand{\sn}{SN\xspace}
\newcommand{\sne}{SNe\xspace}
\newcommand{\snii}{SN~II\xspace}
\newcommand{\sneii}{SNe~II\xspace}

\newcommand{\microns}{$\mu$m\xspace}

\newcommand{\Msun}{\ensuremath{\text{M}_{\odot}}\xspace}
\newcommand{\Rsun}{\ensuremath{\text{R}_{\odot}}\xspace}

\newcommand{\mdotunits}{\Msun~yr$^{-1}$\xspace}
\newcommand{\acko}{SN~2022acko\xspace}
\newcommand{\ixf}{SN~2023ixf\xspace}
\newcommand{\ggi}{SN~2024ggi\xspace}
\newcommand{\vmax}{\ensuremath{v_{\text{max}}}\xspace}
\newcommand{\vphot}{\ensuremath{v_{\text{phot}}}\xspace}

\newcommand{\jwst}{\textit{JWST}\xspace}

\newcommand{\hl}[2]{{#1}\ensuremath{{#2}}\xspace}

\makeatletter
\def\parse#1#2#3#4#5{\@parse#4\@nil}

\def\@parse#1.#2\@nil{
  \def\current@type{#1}
  \ifdefstring{\current@type}{figure}{\@@parse#2\@nil}
    {\ifdefstring{\current@type}{table}{\@@parse#2\@nil}}
    {}
  }
\def\@@parse#1.#2\@nil{\def\current@number{#2}}

\graphicspath{{./figs/}}

\begin{document}

\title{JWST Observations of SN 2024ggi I:
  Interpretation and Model Comparison of the Type II Supernova 2024ggi at 55 days Past Explosion}

\newcommand{\PSI}{\affiliation{Planetary Science Institute, 1700 East Fort
  Lowell Road, Suite 106,Tucson, AZ 85719-2395 USA}}
\newcommand{\HS}{\affiliation{Hamburger Sternwarte, Gojenbergsweg 112, 21029 Hamburg, Germany}}
\newcommand{\IFA}{\affiliation{Institute for Astronomy, University of Hawai’i at Manoa, 2680 Woodlawn Dr., Hawai’i, HI 96822, USA}}
\newcommand{\VT}{\affiliation{Department of Physics, Virginia Tech,
    850 West Campus  Drive, Blacksburg VA, 24061, USA}}
\newcommand{\GRFP}{\altaffiliation{National Science Foundation Graduate Research Fellow}}

\newcommand{\FINESST}{\altaffiliation{NASA FINESST Future Investigator}}

\newcommand{\STSci}{\affiliation{Space Telescope Science Institute, 3700 San Martin Drive, Baltimore, MD 21218-2410, USA}}
\newcommand{\FSU}{\affiliation{Department of Physics, Florida State
    University, Tallahassee, FL 32306, USA}}
\newcommand{\Carnegie}{\affiliation{Observatories of the Carnegie
    Institution for Science, 813 Santa Barbara St., Pasadena, CA 91101, USA}}
\newcommand{\MSU}{\affiliation{Department of Physics \& Astronomy,
    Michigan State University, East Lansing, MI, USA}}
\newcommand{\TAMU}{\affiliation{George P. and Cynthia Woods Mitchell
    Institute for Fundamental Physics and Astronomy,
    Department of Physics and Astronomy, Texas 
             A\&M University, College Station, TX 77843, USA}}
\newcommand{\IALP}{\affiliation{Instituto de Astrof\'isica de La Plata
    (IALP), CONICET, Paseo del Bosque S/N, B1900FWA La Plata, Argentina}}
\newcommand{\LaPlata}{\affiliation{Facultad de Ciencias Astron\'omicas
    y Geof\'isicas Universidad Nacional de La Plata, Paseo del Bosque,
    B1900FWA, La Plata, Argentina}}
\newcommand{\WPI}{\affiliation{Kavli Institute for the Physics and
    Mathematics of the Universe (WPI), The University of Tokyo,
    Kashiwa, 277-8583 Chiba, Japan}} 

\newcommand{\ICE}{\affiliation{Institute of Space Sciences (ICE,
    CSIC), Campus UAB, Carrer de Can Magrans, s/n, E-08193 Barcelona, Spain}}

\newcommand{\IEEC}{\affiliation{Institut d’Estudis Espacials de
    Catalunya (IEEC), E-08034  Barcelona, Spain}} 

\newcommand{\LCO}{\affiliation{Las Campanas Observatory, Carnegie
    Observatories, Casilla 601, La Serena, Chile}} 

\newcommand{\Aarhus}{\affiliation{Department of Physics and Astronomy,
    Aarhus University, Ny  Munkegade 120, DK-8000 Aarhus C, Denmark.}} 

\newcommand{\OU}{\affiliation{Homer L.~Dodge Department of Physics and
  Astronomy, University of Oklahoma, 440 W. Brooks, Rm 100, Norman, OK
  73019-2061}}  

\newcommand{\UCSC}{\affiliation{Department of Astronomy and Astrophysics,
  University of California, Santa Cruz, CA 95064, USA}} 
\newcommand{\Melbourne}{\affiliation{School of Physics, The University of
  Melbourne, VIC 3010, Australia}}

\newcommand{\LPNHE}{\affiliation{LPNHE, (CNRS/IN2P3, Sorbonne
  Universit\'e, Universit\'e Paris Cit\'e), Laboratoire de Physique
  Nucl\'eaire et de Hautes \'Energies, 75005, Paris, France}}

\newcommand{\Princeton}{\affiliation{Princeton University, 4 Ivy Lane,
    Princeton, NJ 08544, USA}}

\newcommand{\Berkeley}{\affiliation{Department of Astronomy,
    University of California, Berkeley, CA 94720-3411, USA}}

\newcommand{\Tsinghua}{\affiliation{Physics Department, Tsinghua
    University, Beijing, 100084, China}}

\newcommand{\Thailand}{\affiliation{National Astronomical Research
    Institute of Thailand, 260 Moo 4, Donkaew, Maerim, Chiang Mai,
    50180, Thailand}}

\newcommand{\UVA}{\affiliation{Department of Astronomy, University of
    Virginia, 530 McCormick Rd, Charlottesville, VA 22904, USA}}

\newcommand{\LJMU}{\affiliation{Astrophysics Research Institute,
    Liverpool John Moores University, 146 Brownlow Hill, Liverpool L3
    5RF, UK}}

\newcommand{\MPIA}{\affiliation{Max-Planck-Institut f\"ur Astrophysik,
    Karl-Schwarzschild Stra{\ss}e 1, 85748 Garching, Germany}}

\newcommand{\JHU}{\affiliation{Physics and Astronomy Department,
    Johns Hopkins University, Baltimore, MD 21218, USA}}

\newcommand{\OSU}{\affiliation{Department of Astronomy, The Ohio State
    University, Columbus, OH, USA}}

\newcommand{\CCAP}{\affiliation{Center for Cosmology and Astroparticle
    Physics, The Ohio State University, Columbus, OH, USA}}

\newcommand{\MIT}{\affiliation{Department of Physics and Kavli Institute for Astrophysics and Space Research, Massachusetts Institute of Technology, 77 Massachusetts Avenue, Cambridge, MA 02139, USA}}

\newcommand{\nextinstitute}{\affiliation{Put the institute of the new author here}}

\author[0000-0001-5393-1608]{E.~Baron}
\email{ebaron@psi.edu}
\PSI
\HS

\author[0000-0002-5221-7557]{C. Ashall}
\email{cashall@hawaii.edu}
\IFA
\VT

\author[0000-0002-7566-6080]{J. M. DerKacy}
\email{jderkacy@stsci.edu}
\STSci
\VT

\author[0000-0002-4338-6586]{P.~Hoeflich}
\email{phoeflich77@gmail.com}
\FSU

\author[0000-0001-7186-105X]{K. Medler}
\email{kyle.medler@sky.com}
\IFA
\VT

\author[0000-0002-9301-5302]{M.~Shahbandeh}
\email{mshahbandeh@stsci.edu}
\STSci

\author[0009-0001-9148-8421]{E.~Fereidouni}
\email{ef22g@fsu.edu}
\FSU

\author[0000-0002-7305-8321]{C.~M.~Pfeffer}
\GRFP
\IFA
\VT
\email{cmpfeffer@vt.edu}

\author[0000-0001-5888-2542]{T.~Mera}
\email{tycomera@gmail.com}
\FSU

\author[0000-0003-3953-9532]{W.~B.~Hoogendam}
\GRFP
\IFA
\email{willemh@hawaii.edu}

\author[0000-0001-6107-0887]{S. Shiber}
\FSU
\email{sshiber1@lsu.edu}

\author[0000-0002-4449-9152]{K. Auchettl}
\email{katie.auchettl@unimelb.edu.au}
\Melbourne
\UCSC

\author[0000-0001-6272-5507]{P. J.~Brown}
\email{grbpeter@yahoo.com}
\TAMU

\author[0000-0003-4625-6629]{C.~R.~Burns}
\email{cburns@carnegiescience.edu}
\Carnegie

\author[0000-0002-5380-0816]{A. Burrow}
\email{anthony.r.burrow-1@ou.edu}
\OU
  
\author[0000-0003-4263-2228]{D.~.A.~Coulter}
\email{dcoulter@stsci.edu}
\JHU
\STSci

\author[0000-0003-0209-674X]{M. Engesser}
\email{mengesser@stsci.edu}
\STSci

\author[0000-0001-5247-1486]{G.~Folatelli}
\email{gaston@fcaglp.unlp.edu.ar}
\IALP
\LaPlata
\WPI

\author[0000-0003-2238-1572]{O. Fox}
\email{ofox@stsci.edu}
\STSci

\author[0000-0002-1296-6887]{L. Galbany}
\email{lluisgalbany@gmail.com}
\ICE
\IEEC

\author[0000-0002-5063-0751]{M. Guolo}
\email{mguolop1@jhu.edu}
\JHU

\author[0000-0001-9668-2920]{J.~T.~Hinkle}
\FINESST
\IFA
\email{jhinkle6@hawaii.edu}

\author[0000-0003-1059-9603]{Mark~E.~Huber}
\IFA
\email{mehuber7@hawaii.edu}

\author[0000-0003-1039-2928]{E.~Y.~Hsiao}
\email{yichi.hsiao@gmail.com}
\FSU

\author[0000-0001-6069-1139]{T. de Jaeger}
\email{dejaeger.thomas@gmail.com}
\LPNHE

\author[0000-0002-6230-0151]{D.~O.~Jones}
\IFA
\email{dojones@hawaii.edu}

\author[0000-0001-8367-7591]{S. Kumar}
\email{sahanak@gmail.com}
\UVA

\author[0000-0002-3900-1452]{J. Lu}
\email{lujingeve158@gmail.com}
\MSU

\author[[0000-0001-6876-8284]{P. A. Mazzali}
\email{P.Mazzali@ljmu.ac.uk}
\LJMU
\MPIA

\author[0000-0002-4338-6586]{N.~Morrell}
\email{nmorrell@carnegiescience.edu}
\LCO

\author[0000-0003-2734-0796]{M.~M.~Phillips}
\email{mmp@lco.cl}
\LCO

\author[[0000-0002-4410-5387]{A. Rest}
\email{arest@stsci.edu}
\STSci
\JHU
 
\author[0000-0002-8102-181X]{N. B. Suntzeff}
\email{nsuntzeff@tamu.edu}
\TAMU

\author[0000-0003-4631-1149]{B.~J.~Shappee}
\email{shappee@hawaii.edu}
\IFA

\author[0009-0008-3724-1824]{Jennifer~Shi}
\Melbourne
\email{jennifer.shi@student.unimelb.edu.au}

\author[0000-0002-5571-1833]{M.~D.~Stritzinger}
\email{max@phys.au.dk}
\Aarhus

\author[0000-0002-7756-4440]{L. Strolger}
\email{strolger@stsci.edu}
\STSci

\author[0000-0001-7380-3144]{T. Temim}
\email{temim@astro.princeton.edu}
\Princeton

\author[0000-0002-1481-4676]{S. Tinyanont}
\email{samaporn@narit.or.th}
\Thailand

\author[0000-0002-2471-8442]{M. Tucker}
\email{tuckerma95@gmail.com}
\CCAP
\OSU

\author[0000-0001-7092-9374]{L. Wang}
\email{lifan@tamu.edu}
\TAMU

\author[0000-0001-5233-6989]{Q. Wang}
\email{qnwang@mit.edu}
\MIT

\author[0000-0002-6535-8500]{Y. Yang}
\email{yiyangtamu@gmail.com}
\Tsinghua
\Berkeley

\submitjournal{ApJ}

\received{\today}
\revised{\today}
\accepted{\today}

\correspondingauthor{Eddie Baron}
\email{ebaron@psi.edu}

\begin{abstract}
We present panchromatic $0.4$--$21$~\microns observations of the
nearby ($\sim 7.2$~Mpc) Type~II supernova 2024ggi, obtained during the
plateau phase at $\sim55$~d past explosion. Our dataset includes
\jwst spectra spanning $1.7$--$14$~\microns, MIR imaging at 7.7 and
21~\microns, and near-simultaneous ground-based optical and  NIR
spectra covering $0.32-1.8$~\microns. The NIR and MIR
spectral features of \ggi are dominated by \ion{H}{1} emission.
We present line IDs and a toy PHOENIX/1D model that reproduces the observations well, especially the continuum redward of 0.9~\microns
We compare \ggi to SN~2022acko and SN~2023ixf, two other Type~II
supernovae  that were also
observed by \jwst, and highlight key similarities and differences in
their spectral features. 
No evidence for a MIR excess or dust is found  at these epochs, with the
model matching the observed flux out to 21~\microns.  We discuss the
model's shortcomings, focusing on  the density profile, which suppresses line
blanketing and produces features in the optical that are too
narrow. Our results show the power of panchromatic 
studies in both exploring the nature of 
the SN ejecta and constraining detailed models of SNe.

 \end{abstract}

\section{Introduction}
\label{sec:introduction}

Hydrogen-rich core-collapse supernovae (CC~SNe), classified as Type II
supernovae (SNe II), mark the deaths of
massive stars with initial masses $\gtrsim 8$~\Msun. These progenitors
are typically red supergiants, although exceptions exist, such as the
blue supergiant progenitor of SN~1987A
\citep[e.g.,][]{Arnett:1989}. Binary evolution likely plays a
significant role in shaping the final evolutionary state of the
progenitor. Massive stars --- at the end of their lives --- form
degenerate iron cores which, upon exceeding the Chandrasekhar mass,
undergo gravitational collapse to nuclear densities. This process
releases $\sim100$~foe of energy \citep[1 foe = $10^{51}$~ergs ][]{Brown:2005}, primarily in the form of neutrinos, but also powers the optical supernova display. The explosion is inherently aspherical, and fully modeling the collapse and explosion remains computationally challenging. As such, the detailed mechanisms driving core-collapse and the resulting observable signatures remain  active areas of research \citep[e.g.,][]{Janka:2025,Vartanyan:2025}.

To date, the number of panchromatic optical to mid-infrared
observations of SNe II is limited, because much of the 
wavelength regime is not observable from the ground. Yet, the
near-infrared (NIR) and mid-infrared (MIR) contain a wealth of
information because lines are less blended and more varied compared to
the optical and ultraviolet. The lower optical depths in the NIR and
MIR also reveal the inner ejecta at earlier times \citep{Meikle:1993} and at
later times can contain signatures of both molecule and dust formation
\citep{Spyromilio:1988,Wooden:1993,Spyromilio:2001,Rho:2018}.

Infrared (IR) observations of CC~SNe were pioneered with the explosion of  
SN~1987A \citep{Bouchet:1987,Oliva:1987,
Catchpole:1988,Elias:1988,Rank:1988,
Whitelock:1988,Catchpole:1989,Meikle:1989,Sharp:1990,
Bouchet:1991,Danziger:1991, Meikle:1991,Meikle:1993}.
These studies demonstrated the wealth of information available at
these wavelengths and laid the groundwork for the interpretation of future IR observations.
SN~1987A continues to be studied in the NIR and MIR with the James
Webb Space Telescope (\jwst) \citep{Larsson:2023,Jones:2023}. 

Since  SN~1987A, many SNe~II have been observed using ground-based NIR (1-2~$\micron$) resources 
 allowing the characterization of spectral line evolution from a couple of days past the explosion through the nebular phase
\citep{Benetti:2001,Hamuy:2001,Elmhamdi:2003,Pozzo:2006,Pastorello:2009,Maguire:2010,Fraser:2011,Tomasella:2013,DallOra:2014,Jerkstrand:2014,Morokuma:2014,Takats:2014,Takats:2015,Valenti:2015,Valenti:2016,Rho:2018,Bostroem:2019,Szalai:2019,Tucker:2024}.
For example, SN~1999em was well observed with NIR spectroscopy,
providing line identifications and additional analysis
\citep{Hamuy:2001,Elmhamdi:2003}. 
\citet{Davis:2019} presented the results of 30 SNe~II with 81 NIR
spectra obtained by the Carnegie Supernova Project 2 \citep[CSP2, ][]{Phillips:2019,Hsiao:2019} and were
able to discern correlations between the strength of the \ion{He}{1}
lines and the formation of CO. However, at wavelengths longer than $2$~\microns there is still a dearth of data.

The era of MIR spectroscopy ($\lambda>2$~\microns) for CC-SNe began to take traction with the launch of the Spitzer Space Telescope
(\textit{SST}) with observations of SNe 2004et, 2004dj, and 2005af \citep{Kotak:2009,Fox:2011,Szalai:2013}, which showed the evolution
of CO, and dust in the ejecta.  However, these datasets did not have
continuous wavelength coverage from the optical to MIR, and hence
lacked complete coverage of key spectral features such as the CO
fundamental band at
$\sim 4$~\microns.

More recently, a number of CC~SNe  have been observed  with
\jwst, including the SNe II 2022acko \citep{Shahbandeh:2025}, and
2023ixf \citep{DerKacy:2025a,Medler:2025b}. 
The launch of \jwst opened the window to continuous spectral coverage
into the MIR, allowing the construction of  panchromatic multi-epoch
optical to MIR data sets of SNe. Observations of SN~2022acko at $\sim
50$~d show many series of H-lines and a lack of pre-existing molecules
or dust \citep{Shahbandeh:2025} revealing the promise that the
construction of 
these datasets offer.
SNe II are heterogeneous at optical wavelengths \citep{Hinds:2025}, 
due to variations in  both the progenitor systems and explosion
processes. IR observations probe deeper into the ejecta than optical
photons, opening a new window into the ejecta
structure.

\ggi was  discovered by the Asteroid Terrestrial-impact
Last Alert System 
 \citep[ATLAS][]{Srivastav_etal_2024,Tonry_etal_2024,Chen:2024b} on
2024-04-11 (MJD 60411.14) in the host galaxy NGC 3621 located at a distance of
$7.24 \pm 0.20$~Mpc \citep{Saha_2006} --- making \ggi, along with SN~2023ixf, one of the closest SNe in modern times.  The transient was
classified  as a SN II with flash ionization features
\citep[][]{Hoogendam_etal_2024,Zhai_etal_2024}.  Due to the
proximity of \ggi and its early discovery, \jwst
Director Discretionary Time (DDT)  was approved to observe it
throughout its evolution through program JWST-GO-6677/6716
\citep[PI: Ashall;][]{Ashall_2024_ggi_b,Ashall_2024_ggi_c}\footnote{\url{https://chrisashall.com/mid-infrared-supernova-collaboration-mirsnac/}}.  The purpose of this program is to follow the NIR and MIR spectroscopic
evolution of \ggi from the plateau phase to $\sim$300~d past
explosion.

We present  the first epoch of these observations, which were obtained
$\sim$55~d post explosion. In addition to the spectra and photometry
obtained with  \jwst, optical and NIR spectra were obtained with
Spectroscopic Classification of Astronomical Transients (SCAT) Program \citep{Tucker22} and The Hawaii Infrared
Supernova Study  \citep[HISS; ][]{Medler:2025a}, respectively.
The features in the
observed spectra are  identified. We 
present  NLTE \phx models that reproduce the features and
continuum in NIR and MIR and discuss their shortcomings. 

\subsection{Previous Work}

\label{sec:others}

While we focus on our panchromatic observations at one epoch on the plateau, \ggi has been well observed at early times. Here  we briefly review the previous work on \ggi.

High resolution spectra of \ggi were obtained with the MIKE
spectrograph on the Magellan Clay telescope 21, 27, 28, and 34 hours
after discovery \citep{Shrestha_etal_2024,Pessi_etal_2024}. They
show solid evidence of flash ionization features. The early spectra
were compared to detailed radiative transfer models and a mass loss
rate of $\dot M = 4\times 10^{-3}$~\mdotunits was inferred
\citep{Shrestha_etal_2024}. Thus, there is evidence for
  interaction of the ejecta with the surrounding circumstellar medium
  (CSM).

Early-time (+8---+17~d) observations using ALMA millimeter data 
\citep{Hu:2024} and high-cadence optical spectroscopy
\citep{Zhang:2024,Pessi_etal_2024} have been crucial in
probing the CSM around the progenitor. While
ALMA did not detect emission, the results help distinguish between a
steady wind and an eruptive mass-loss scenario, with an eruptive model
favored and a derived mass-loss rate of approximately $5 \times
10^{-3}$~\mdotunits \citep{Hu:2024}. Early light curves are well
described by a broken power law with a break at $\sim1$~day
\citep{Shrestha_etal_2024}. Flash ionization features indicate
CSM velocities of $\sim 20$–$40$~\kmps, radii of $3–6\times
10^{14}$~cm, and mass-loss rates ranging from $10^{-3}$ to
$10^{-1}$~\mdotunits
\citep{Shrestha_etal_2024,Jacobson-Galan_etal_2024,Chen:2024b,Zhang:2024}.

The CSM mass is estimated at $0.4$\,\Msun
\citep{Chen:2024b}, and the maximum observed velocities reach up to
$8000$~\kmps \citep{Zhang:2024}.  Spectroscopic observations within
the first two days post-discovery reveal flash ionization features of
\ion{H}{1}, \ion{He}{1}, \ion{He}{2}, \ion{C}{3}, \ion{C}{4}, \ion{N}{3}, and \ion{Si}{4}, 
\citep{Pessi_etal_2024,Chen:2024b}. The line profiles are best fit with Voigt
profiles, and the broad emission components increase in velocity over
time, suggesting an accelerated, radiatively driven wind
\citep{Pessi_etal_2024}. Notably, high-ionization lines such as \ion{He}{2}
are sometimes absent \citep{Zhang:2024}. Comparisons to CMFGEN \citep{Hillier_Miller:1998} and
other detailed models reproduce the ionization state at $\sim1$~day
and the blue edge of \hl{H}{\alpha} at $\sim15$~d, but no model fits all
features perfectly \citep{Jacobson-Galan_etal_2024,Zhang:2024}.

Light curve modeling, including the rapid early
blueward evolution and the presence of flash ionization features,
supports a scenario involving wind shock breakout
\citep{Chen:2024b}. Explosion energies are inferred to be
$2\times10^{51}$~erg, with ejecta masses of $1.2$~\Msun and CSM radii of
$4–6\times10^{14}$~cm
\citep{Chen:2024a,Chen:2024b,Jacobson-Galan_etal_2024,Zhang:2024}.
The derived
mass-loss rates and CSM masses are difficult to reconcile with
steady-state wind mass loss \citep{Smith:2017} or standard neutrino-driven core-collapse
explosion mechanisms \citep{Soker:2024}.

Progenitor properties have
been constrained using pre-explosion imaging from the Hubble Space
Telescope (\emph{HST}), the Spitzer Space Telescope (SST), and
the Dark Energy Camera Legacy Survey \citep{Xiang:2024,Chen:2024a}.
The progenitor is identified as a variable star
consistent with a red supergiant, with a mass-loss rate of less than
$3\times10^{-6}$~\mdotunits prior to the enhanced mass loss phase
inferred from the flash ionization
\citep{Xiang:2024}. This result is corroborated by light curve and
spectral modeling  
that  finds that the CSM distribution likely consists of a two-component
structure with a compact core and extended tail
\citep{Ertini:2025}.  \citet{Ertini:2025} find the 
progenitor fits
a zero-age main  $M_\text{ZAMS} = 15$~\Msun, with a  pre-SN mass
14.1~\Msun, and  a  mass-loss rate of $4\times
10^{-3}$~\mdotunits,  giving   an
explosion energy  1.3 foe.
Analysis of the
progenitor environment indicates a uniform spatial distribution and a
star formation history corresponding to a progenitor mass of
$10.2$~\Msun \citep{Hong:2024}. Semi-analytic modeling and archival
imaging yield ejecta masses of $1.2$~\Msun, a progenitor radius of
$555$~\Rsun, and progenitor masses in the range $14–17$~\Msun
\citep{Chen:2024a}.

\begin{figure*}[ht]
    \centering
    \includegraphics[width=\textwidth]{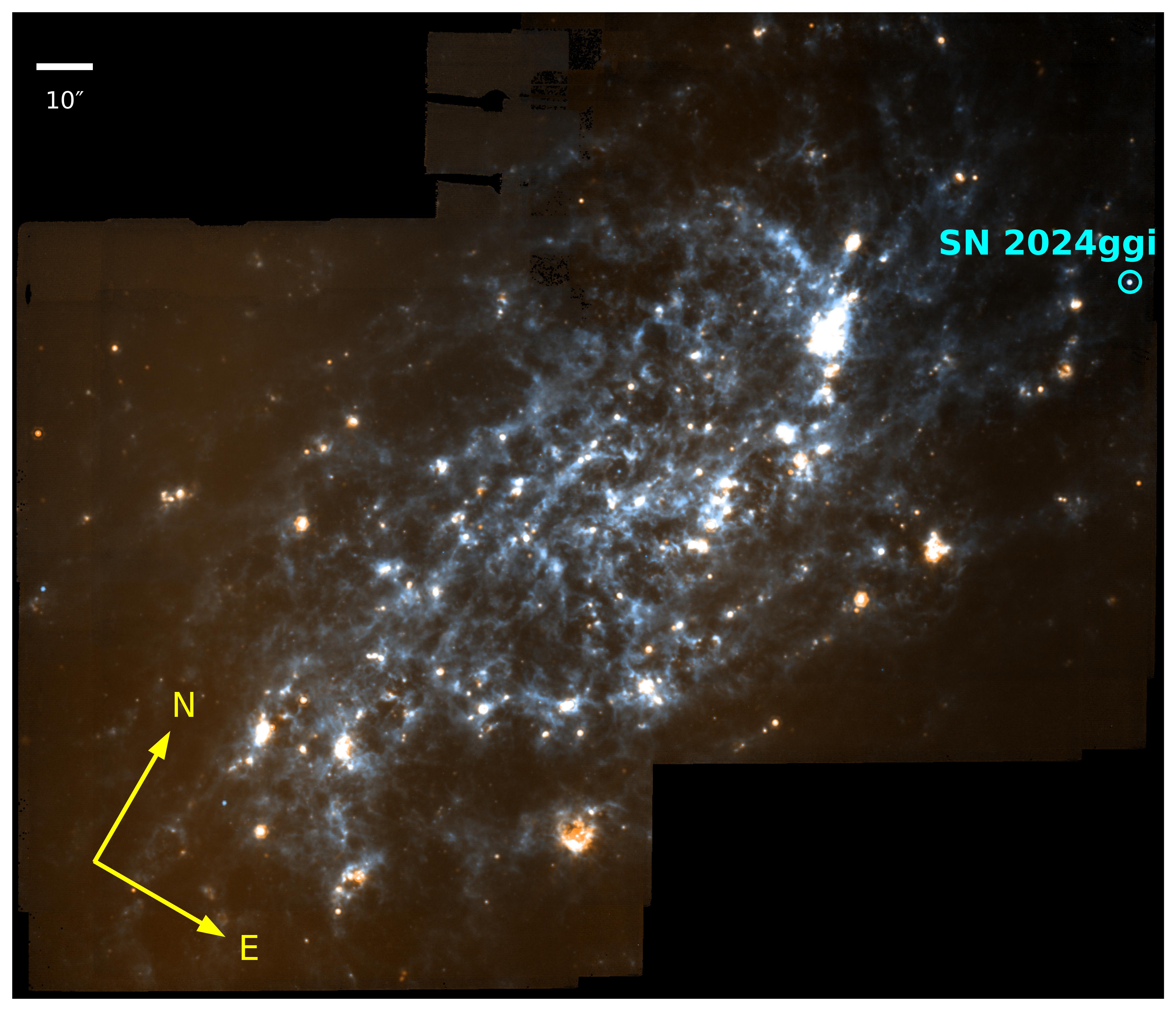}
    \caption{A composite two-band color image constructed from the \textit{MIRI} imaging of NGC 3621 using filters \textit{F700W} and \textit{F2100W}. The location of \ggi is highlighted in the turquoise circle. }
    \label{fig:colorimage}
\end{figure*}

\section{Observations and Data Reduction}
  \label{sec:data}

  \begin{table}
    \centering
    \caption{Basic observational parameters of \ggi used in this work \label{tab:ggi_details}}
    \begin{tabular}{ccc}
    \hline
    \hline
    Parameter & Value & Source \\
    \hline
    RA & 11$^{h}$18$^{m}$22$^{s}$.087 & (1,2)  \\
    DEC & $-$32\arcdeg 50 \arcmin 15 \arcsec.27 & (1,2) \\ 
    $T_\text{exp}$ (MJD) & $60410.80 \pm 0.34$ & (3) \\
    $M_{V_\text{max}}$ (mag) & $\sim-17.72$ & (4) \\
    $z$ & 0.002435 & (5) \\
    Distance (Mpc) & $7.24 \pm 0.20$ & (5) \\
    $\mu$ (Mpc) & $ 29.30\pm 0.03 $ & (5) \\ 
    $E$(B-V)$_\text{MW}$ (mag) & $0.054\pm 0.02$ & (4,6) \\
    $E$(B-V)$_\text{Host}$ (mag)& $0.034 \pm 0.02$ & (5) \\
    $E$(B-V)$_\text{Cloud}$ (mag)& $0.066 \pm 0.02$ & (4) \\
    Host & NGC 3621 & (1)  \\
    \hline
    \end{tabular}
    \tablerefs{(1) \href{TNS}{https://www.wis-tns.org/object/2024ggi},
      (2) \cite{Chambers_2024_SN24ggi},
      (3) \cite{Jacobson-Galan_etal_2024}, (4) \cite{Shrestha_etal_2024}, (5) \cite{Saha_2006}, (6) \cite{Schlafly_2011}}
\end{table}

The basic parameters  we use throughout this work are presented in  \autoref{tab:ggi_details}. We take the time of first light from \citet{Jacobson-Galan_etal_2024}:
MJD $60410.80 \pm 0.34$~d, which is
based on the average between the times of last non-detection
and first detection.  (Unless stated otherwise, we will quote epochs
in time since first light, referring interchangeably to this as time
since explosion). We adopt the total extinction estimate of $E(B-V)_\text{total}$= 0.154~mag
 \citep{Shrestha_etal_2024}, which is similar to values found in other
investigations  \citep{Jacobson-Galan_etal_2024,Pessi_etal_2024}. We
note that in addition to extinction in the host and the Galaxy,
\citet{Shrestha_etal_2024} include extinction from an intervening cloud
at redshift $z=0.00039$, and that their extinction estimates rely on
the Na D equivalent width formula of \citet{Poznanski_2012}, which are
known to have a large error \citep{Phillips_2013}.

\label{sec:preprocessing}

\subsection{JWST Spectra}

 Spectroscopic observations were obtained at $\sim$55~d past explosion
 using the fixed slit with the Near Infrared Spectrograph (NIRSpec)
 and the  Mid-Infrared Instrument (MIRI) low-resolution spectrometer
 (LRS). This produced continuous spectral data from
 $\sim$1.6-14~$\micron$.
The resolving power of these observations is $\sim$1000 for NIRSpec and
$\sim$100 for MIRI/LRS. 
 All data were reduced using the \jwst pipeline version 1.18.0 and CRDS context \texttt{jwst\_1371.pmap}. A log of the spectral observations can be found in \autoref{tab:JWST_spec_info}.

\subsection{JWST Imaging}
The Physics at High Angular resolution in Nearby
GalaxieS (PHANGS) collaboration ( \citep{Lee22,Leroy:2023} obtained serendipitous
imaging of \ggi  in the \jwst MIRI filters
\textit{F770W} and \textit{F2100W} at $\sim 64$~d. As \ggi\ was located at the edge of the field of view, it is only visible in 2 of the 4 dithers in each of these filters.  \autoref{fig:colorimage} presents a color composite of \ggi and the host galaxy NGC 3621.

Point spread function (PSF) photometry was computed using a custom-made python notebook making
use of the  \textit{WebbPSF} photometry package
\citep{Perrin14}\footnote{\url{https://github.com/orifox/psf_phot/blob/main/space_phot/MIRI/miri_1028.ipynb}}. The
PSF width was determined by minimizing the  residual, while also
taking into account the secular increase of  the PSF with wavelength. 
  The fluxes of each dither, where the SN was visible, were averaged, and the error 
  in the flux was obtained using the standard deviation of the fluxes. 
  A log of the photometric observations is presented in  \autoref{tab:JWST_phot}.

\begin{deluxetable}{ccccc}
\tabletypesize{\footnotesize}  \tablewidth{\columnwidth}      \tablecaption{The photometry obtained with \jwst.\label{tab:JWST_phot}}
\tablehead{
\colhead{Filter} & \colhead{Obs. date} & \colhead{Exp. time} & \colhead{Phase} & \colhead{Magnitude} \\
\colhead{}       & \colhead{(MJD)}     & \colhead{(s)}        & \colhead{(days)} & \colhead{(Mag)}
}
\startdata
\textit{F770W}   &  60474.7995 & 44 & 63.84 &  14.885 $\pm$ 0.002 \\
\textit{F2100W}  & 60474.864 & 172 & 63.91 & 16.443 $\pm$ 0.017 \\
\enddata
\end{deluxetable}

\begin{figure*}[ht]
    \centering
    \includegraphics[width=\textwidth,trim={1cm 0 3cm 0},clip]{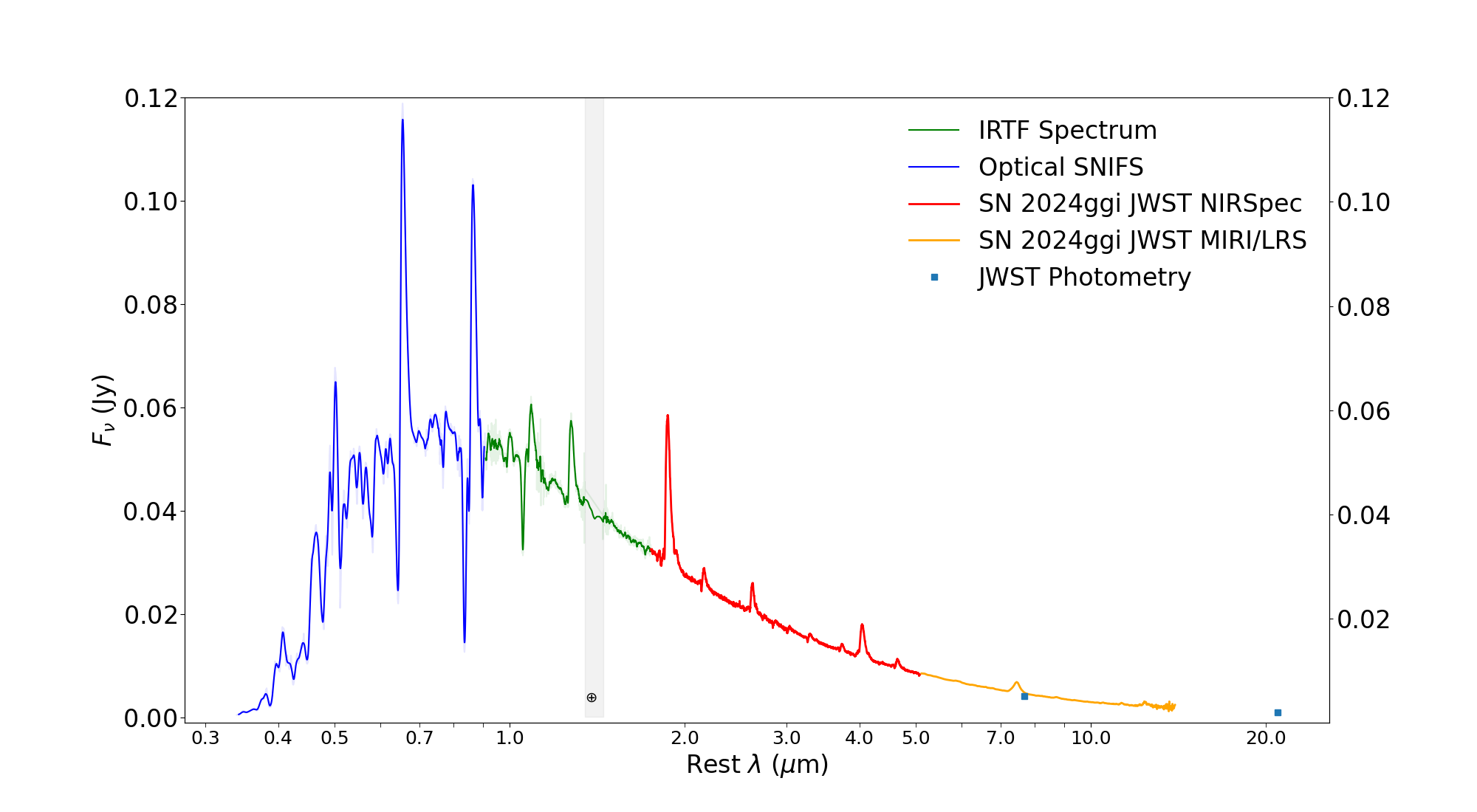}
    \caption{The full dereddened SED of \ggi.
    The UH88  optical spectrum was obtained on MJD
      60470 (day $\sim 60$),  the IRTF  spectrum obtained on day
      44. The NIRSpec spectrum obtained on MJD 60466.46 (day 55)
       and  the MIRI/LRS spectrum obtained
      in the same visit as the NIRSpec.  We note that the \textit{F700W} and
      \textit{F2100W} images were taken $\sim$9~d after the \jwst
      spectra which might explain the flux discrepancy between the
      photometry and the spectra.}  
    \label{fig:fullSED}
\end{figure*}

\subsection{Ground Based Spectra}
In order to fill out the spectral energy distribution (SED), we
obtained optical and NIR spectra at a similar epoch to the \jwst
data. An optical (0.32-0.9~$\micron$) spectrum was obtained as part of
the SCAT program \citep{Tucker22}, using the SuperNova Integral Field Spectrograph \citep[SNIFS;][]{Lantz:2004} mounted on the
University of Hawaii 88-inch telescope. The data were obtained 59.5~d
(MJD=60470.27) past explosion with a 1200~s exposure, and reduced
using the procedure outlined in \citet{Tucker22}. 

To observe the NIR we  also obtained an IRTF spectrum (Program 2024A987, PI: Hoogendam) 44.43~d (MJD=60455.23) past explosion 
as part of  HISS  \citep{Medler:2025a}.
The IRTF spectrum
had an exposure time of  1434.5~s, consisting of 12  A/B individual
119.5~s exposures and was reduced using the methods described by the
HISS team
\citep{Hoogendam:2025a,Hoogendam:2025b,Medler:2025a}.   

The SNIFS spectra are not processed with the full SNFactory pipeline
and are thus  
not spectrophotometric, however, only minor (roughly 10\%) adjustments
in flux were needed to match the IRTF spectrum to that of 
\jwst, which was then matched to the SNIFS spectrum again with minor
adjustments. 
We focus on the flux calibrated  \jwst spectra for the
spectral modeling.

\section{Data Analysis}
\label{sec:data_anal}

\subsection{SED}

\autoref{fig:fullSED} shows the full 0.32-20~\microns SED of \ggi at
$\sim$55~d past explosion. The spectra are consistent with a
recombination-dominated regime characterized by numerous hydrogen line
transitions. Despite the ground-based and \jwst\ spectra being
obtained at a
slightly different phase, they align in flux with only modest matching.  The
photometric data, which was taken $\sim 9$~d after the \jwst data has a flux
slightly below that of the spectra.

\begin{figure*}[ht]
    \centering
    \includegraphics[width=\textwidth]{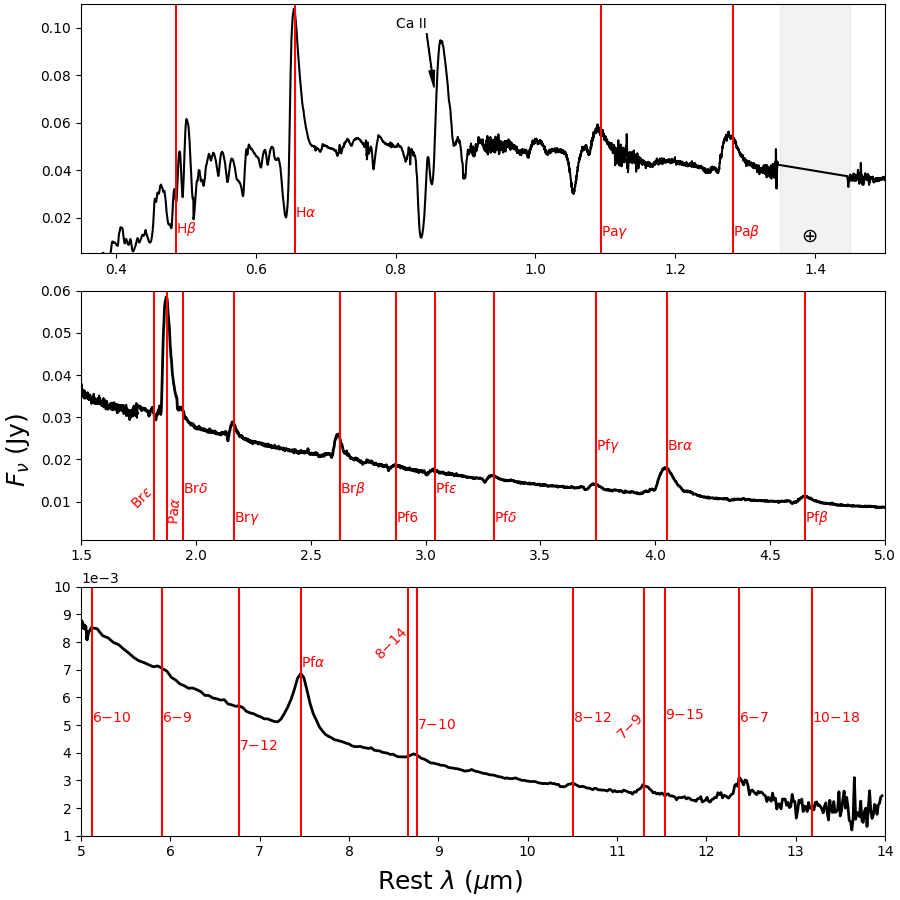}
    \caption{The optical, ground-based NIR, and \jwst \ggi spectra with prominent
      \ion{H}{1} lines 
      identified.}
    \label{fig:h1ids}
\end{figure*}

\autoref{fig:h1ids} shows that, except for the optical wavelengths,   all of the major features in
both the IRTF and the \jwst spectra are accounted for by lines of \ion{H}{1}. The strength and width of these lines reflect transitions within the atomic structure, with the strongest features arising from the most probable or lowest-energy level transitions.

\subsection{Spectral Comparison}

\begin{figure*}[ht]
    \centering
    \includegraphics[width=\textwidth]{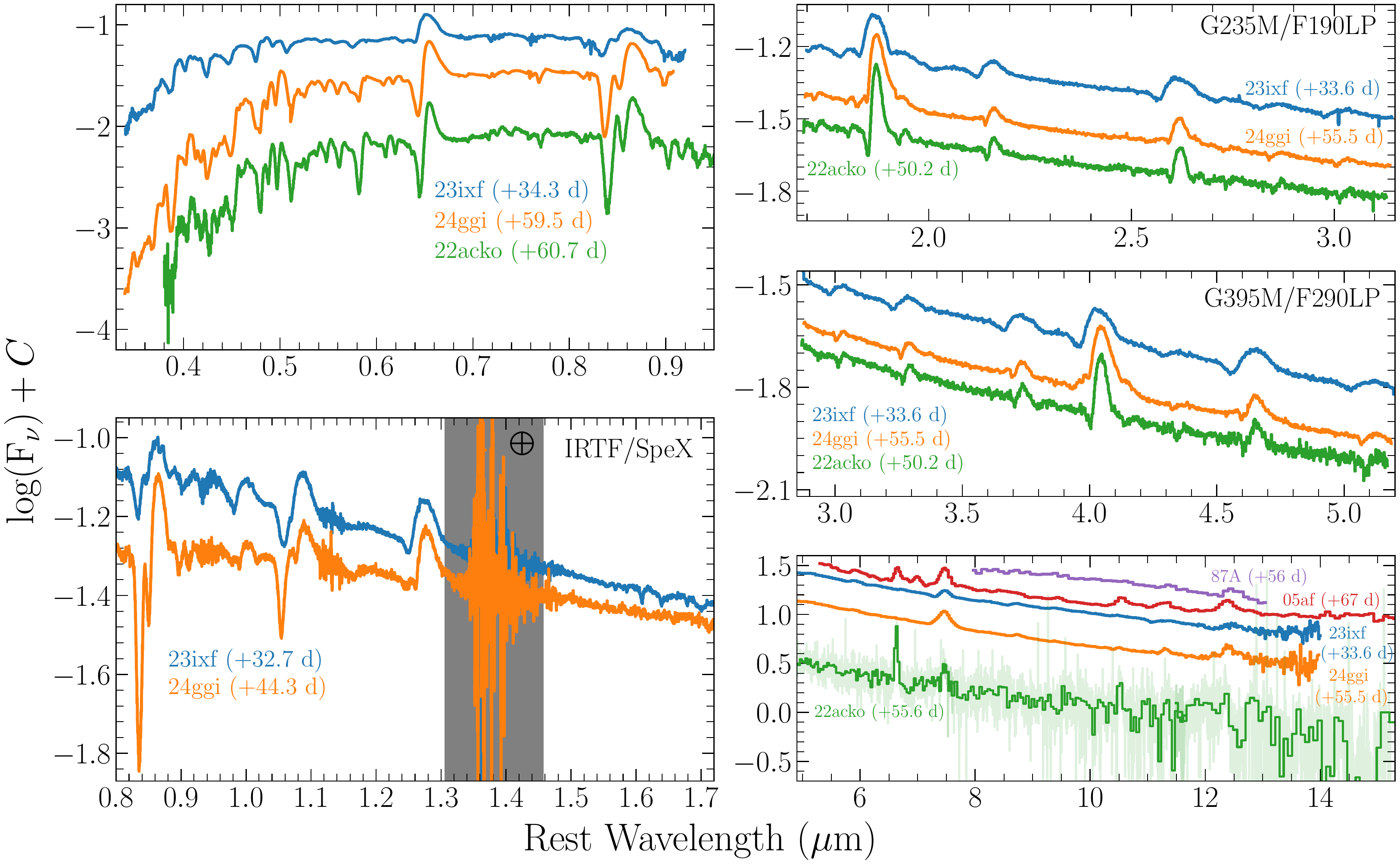}
    \caption{Panchromatic spectral comparison of SNe II observed with JWST 
             (SNe~2022acko, 2023ixf, and 2024ggi) during the plateau phase; 
             and other historical SNe II
             \citep[SNe~1987A][]{Aitken1988} and
             \citep[SN~2005af][]{Kotak2006} observed in  
             the MIR at similar phases. The \acko data shown here have been 
             re-reduced using updated versions of the \jwst pipeline 
             (v1.18.0+\texttt{jwst\_1371.pmap} and v1.17.1+\texttt{jwst\_1363.pmap} 
             for the NIRSpec and MIRI/MRS data respectively) relative to the 
             data first presented in \citet{Shahbandeh:2025}.}
    \label{fig:snecmp}
\end{figure*}

Both \acko \citep{Shahbandeh:2025} and \ixf \citep{DerKacy:2025a} have 
panchromatic datasets (near-) contemporaneous with their \jwst observations
during the plateau phase. These comparisons are shown in \autoref{fig:snecmp}.
\jwst data were obtained approximately at the mid-point of the plateau phase
for each SN. We note that the spectra of \acko used in the figure have been re-reduced 
using the \texttt{AstroBkgInterp}\footnote{https://github.com/brynickson/AstroBkgInterp}
package, which recovers the SN continuum better than the
reduction 
presented by \citet{Shahbandeh:2025}.

In general, all three \sne show features typical of normal \sneii on
the plateau. In 
the optical, the later observation epoch of \acko and \ggi compared to
that of \ixf results in 
stronger blanketing by iron-group lines in the blue optical wavelengths
and more s-process lines observed.
While \citet{DerKacy:2025a} identify the \ion{He}{1} 1.083~\microns line in the ground-based IRTF spectra of \ixf,
we see no strong evidence
for lines of \ion{He}{1} in \ggi.  The lower velocity of the Paschen 
lines in \ggi results in less blending than is seen in \ixf. The
feature blueward of \hl{Pa}{\gamma} is more likely due to the
\ion{C}{1} 1.069\microns line than to the \ion{He}{1} 1.083\microns
line (see \autoref{sec:disc}).

In the JWST/NIRSpec observations, there is near uniformity among all three 
objects in the H lines which are present in the spectrum. At MIR wavelengths 
($\lambda > 5$~\microns), all three \sne are dominated by hydrogen lines of 
the Pfund, Humphreys, and unnamed higher order series. The emission peaks of 
the strong \hl{Pf}{\alpha} plus \hl{Hu}{\beta} blend and the \hl{Hu}{\alpha} 
line are more prominent in \ggi, similar to SN~2005af, than in SNe~1987A or 
2023ixf. The lower S/N of the \acko data makes distinguishing all but the 
strongest hydrogen lines difficult.

\subsection{Spectral Line Velocities}

\begin{figure*}[ht]
    \centering
    \includegraphics[trim=2cm 0.7cm 2cm 3.5cm,clip=True,width=\textwidth]{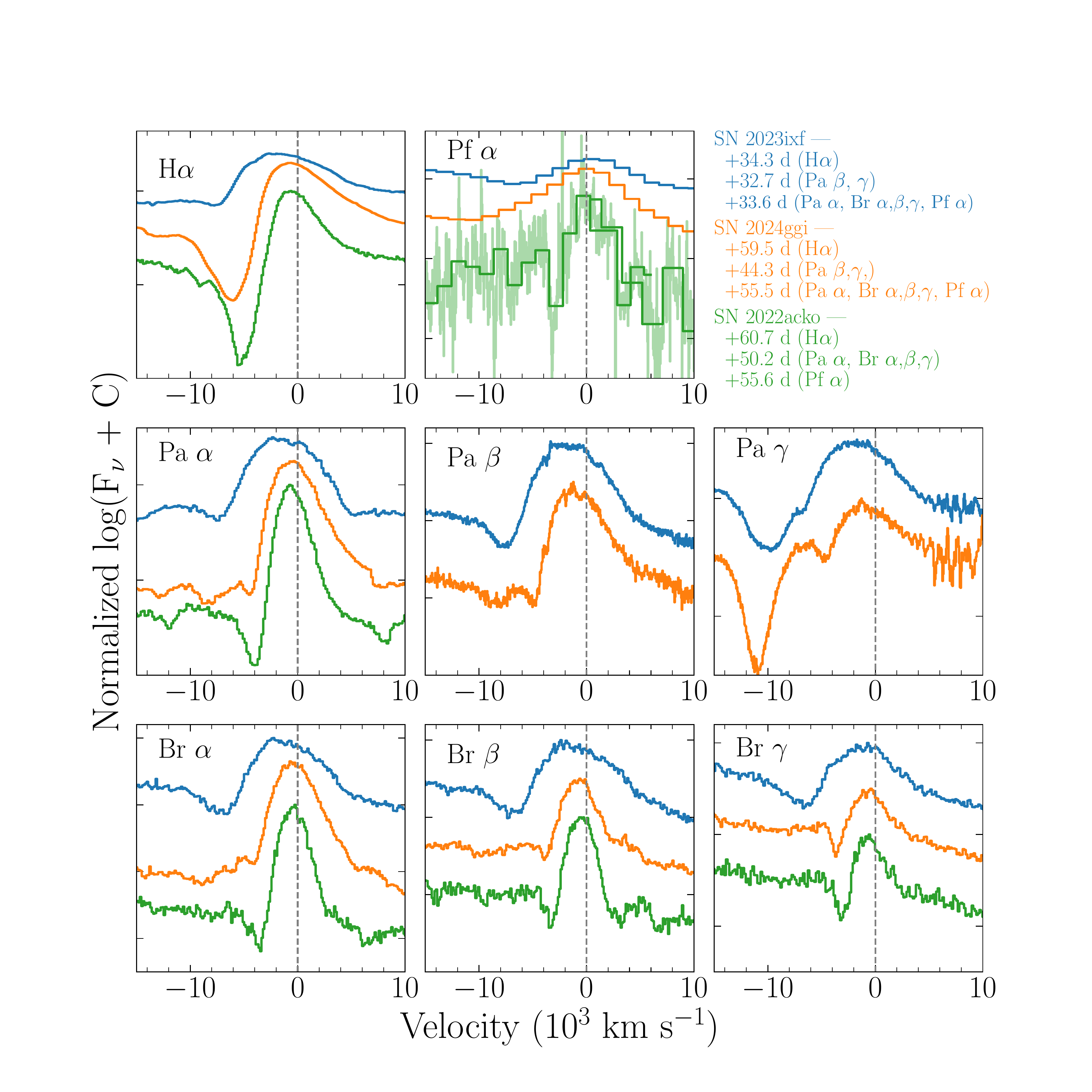}
    \caption{A velocity-space comparison of some strong P-Cygni profiles found in \jwst 
    and ground-based observations of SNe~2022acko, 2023ixf, and 2024ggi.}
    \label{fig:velplot}
\end{figure*}

More detailed comparisons focusing on the hydrogen lines with 
strong P-Cygni features in the three \jwst \sneii (e.g. \hl{H}{\alpha}; 
Pa~$\alpha$, $\beta$, $\gamma$;  Br~$\alpha$, $\beta$, $\gamma$; and 
\hl{Pf}{\alpha}) are shown in \autoref{fig:velplot}. Though the phases 
of the SNe with respect to explosion vary somewhat, the 
wide variation in the line-shapes gives an indication of the degree 
of heterogeneity of the SN~IIP sub-class. One trend that is 
homogeneous across all three \sne is the correlation between line 
velocity (measured from the P-Cygni absorption trough) and the absolute $V$-band magnitude 
\citep{Hamuy:2002,deJaeger:2020}. \acko has  $M_V \sim -15.4$~mag \citep{Bostroem:2023},
while \ixf has $M_V \sim -18.2$~mag \citep{Zheng:2025}, and \ggi has $M_V
\sim -17.7$~mag (\autoref{tab:ggi_details}).
As \autoref{fig:velplot} shows,
this trend is consistent across multiple series of H lines, not
just the \hl{H}{\beta} and \ion{Fe}{2} $\lambda5169$ lines commonly used 
as characteristic measures of the photospheric velocity in \sneii.

Not only are the line profiles heterogeneous across the \snii sample,
but also within each \sn. 
 Unlike \ixf, which shows 
multiple examples of substructure within the H lines 
\citep{Singh:2024,DerKacy:2025a}, \ggi shows no substructures common 
to multiple H series, at this phase. The closest thing to substructure
seen in \ggi  
is a small peak at $\sim -1,300$~\kmps in the Paschen $\beta$, 
$\gamma$, $\delta$, and $\zeta$ lines. The feature may be present 
in other lines, but it is too weak to be distinguished given the line
profiles variability.

\section{Models}

Synthetic spectral models of core-collapse supernovae have a long
history \citep{Payne-Gaposchkin:1940,Branch:1980}, but the advent of
SN~1987A prompted much work
\citep{Lucy:1988,Harkness:1988,Hoeflich:1988,Eastman:1989,Mazzali:1992,Hauschildt:1994,Swartz:1989,Kozma:1998,Mitchell:2002}. Numerical
approaches roughly fall into three classes: i) 
Direct modeling approaches such as that used by the SYNOW code
\citep{Jeffery_1990}; ii) grid based solution of the radiative
transfer equation, such as \phxO \citep{Hauschildt_Baron_1999}, HYDRA
\citep{Hoeflich:1988}, and CMFGEN \citep{Hillier_Miller:1998,Dessart:2005} 
and iii) Monte Carlo methods such as those of Mazzali
\citep{Mazzali:1992}, ARTIS, \citep{Sim:2007,Kromer_Sim:2009}
TARDIS \citep{Kerzendorf_Sim:2014}, SEDONA \citep{Kasen:2006}, and SUMO
\citep{Jerkstrand:2011}.

Here we use SYNOW and \phxO.
SYNOW generates synthetic supernova spectra by modeling the
photospheric phase with a few simplifying assumptions: Spherical
symmetry and homologous expansion; a sharp photosphere emitting  a
black‑body continuum with only line transport which is treated as
resonance scattering in the Sobolev approximation ignoring  ionization
and rate equations. The output is a blackbody continuum overlaid with
blended P-Cygni profiles. We only focus on \hl{H}{\alpha} here, but in
general for each ion the input parameters are:   the Sobolev optical
depth of a reference line photospheric velocity ($\vphot$), and the
e-folding velocity  of the reference line optical depth, $v_e$.
Optical depths decline exponentially with velocity: $\tau(v) =\tau(\vphot)\exp[-(v-\vphot)/v_e]$.
See \citet{Jeffery_1990} and \citet{Branch:2005,Branch:2006} for more details.

\phxO is a generalized stellar atmosphere code that solves the
radiative-transfer equation along characteristic rays in spherical
symmetry, fully accounting for special-relativistic effects. It treats
non-LTE (NLTE) rate equations for a comprehensive set of ions,
Primary
NLTE transitions determine level populations and opacity; weaker lines
are added in LTE using the NLTE ground-state occupation, ensuring no
transition is omitted. 
Energy balance is enforced in the comoving frame, implicitly incorporating
time-dependent effects through PdV work and gamma-ray deposition. The
outer boundary sets the observer-frame bolometric luminosity; the
inner boundary is that the flux is given by diffusion at the 
deepest zone.

\phxO
has been used on a wide variety of supernovae including SN 1993J
\citep{Baron:1993,Baron:1995}, SN~1994I \citep{Baron:1996,Baron:1999},
SN~1998S \citep{Lentz:2001_98S}, SN~1999em
\citep{Baron:2000,Baron:2004}, SN~1987A \citep{Mitchell:2002},
SN~1993W \citep{Baron:2003} and SNe~Ia
\citep{Nugent:1995,Lentz:2001_84A,Lentz:2001_94D,Baron:2006,Bongard:2006,Baron:2008,Bongard:2008,Friesen:2017,DerKacy:2020}.

\subsection{Methods}
\label{sec:methods}

Given that both the ground based and \jwst spectra are dominated by
\ion{H}{1} lines (\autoref{sec:data_anal}), we  model the
spectra to place constraints on the line formation processes in the ejecta. The goal
is to produce a model which matches the overall spectra and not to tweak the model to match every
single feature. 
Thus, using the generalized stellar atmospheres code \phx
\citep{Hauschildt_Baron_1999}, we constructed a simple model assuming a
power-law density profile, 
$\rho \propto v^{-n}$, with $n=13$ and $\vmax = 5900$~\kmps. The model
has solar abundances 
\citep{Asplund_etal_2009}. The low outer velocity is intended to mimic
the  effect of circumstellar interaction. The luminosity is $L =
2\times 10^9 \,\text{L}_\odot$, the ejecta mass (mass exterior to
$\tau_\text{std} = 10$, where $\tau_\text{std}$ is the optical depth
in the continuum at 5000~\mbox{\AA}) is $M_\text{ejecta} = 9.1$~\msol, with kinetic energy 
$E_\text{ejecta} = 0.48$~foe. 
Since the spectra were dominated by hydrogen recombination lines, we
were able to choose rather generic parameters for SNe IIP (based on
previous studies) and thus, did not need to vary our initial parameter
choices, except for the outer pressure (which determines \vmax) in
order to obtain the fiducial model.

\subsection{Results}
\label{sec:results}

\begin{figure*}[ht]
    \centering
    \includegraphics[width=\textwidth]{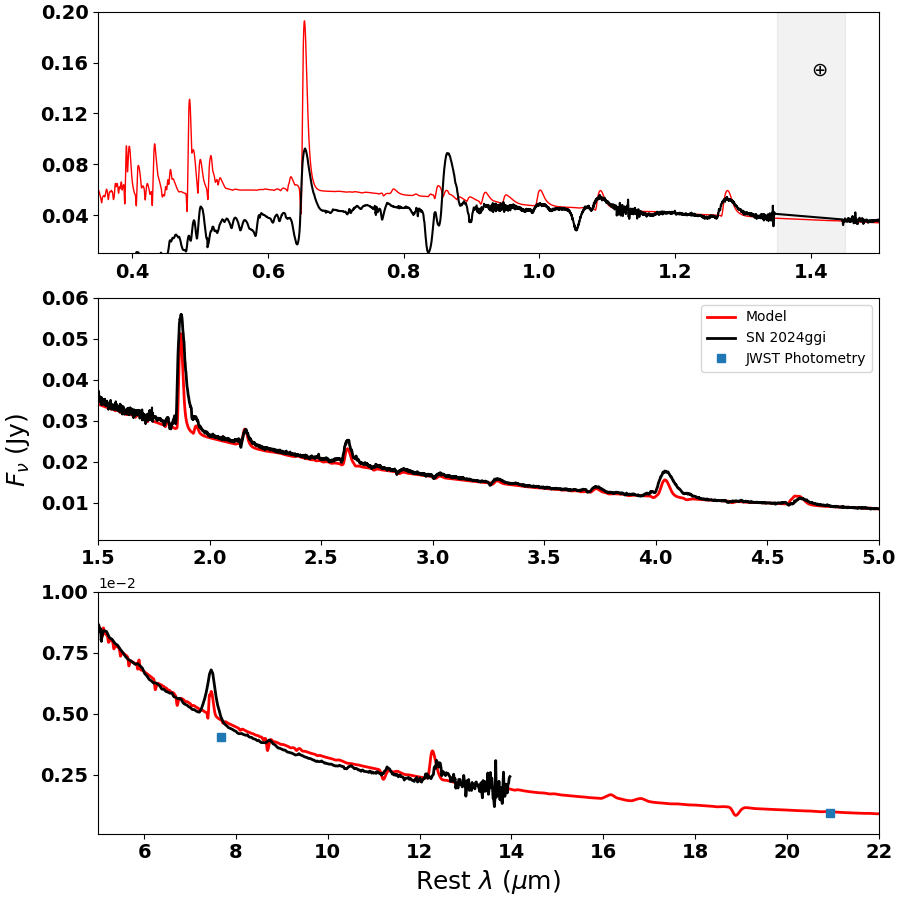}
    \caption{The observed day 55 \ggi ground-based and \jwst spectrum compared to
      the \phx model (see \autoref{sec:results}). In this and
      subsequent model plots the synthetic spectrum has been scaled to
    the observed spectrum at 3~\microns.}
    \label{fig:model_data}
\end{figure*}

\autoref{fig:model_data} shows the fiducial model described above
(\autoref{sec:methods}) compared to the data from the optical to the
MIR. While the \jwst spectra were obtained at 55~d,  the
optical UH88 spectrum  is at
$\sim 60$~d and the IRTF spectrum is at $\sim 44$~d, the model
is constructed at 55~d. The model does a very 
good job of fitting the NIR+MIR continuum ($\lambda >
0.9$~\microns). However, the relatively steep density profile
significantly inhibits line blanketing in the UV, showing that the
true ejecta density profile is not captured by a
single power-law.

Our goal is not to produce a perfect fit to
the full dataset, but we rather focus on the fit associated with the
NIR+MIR data and thus we take
this model as
our fiducial model.

\begin{figure*}[ht]
    \centering
    \includegraphics[width=\textwidth]{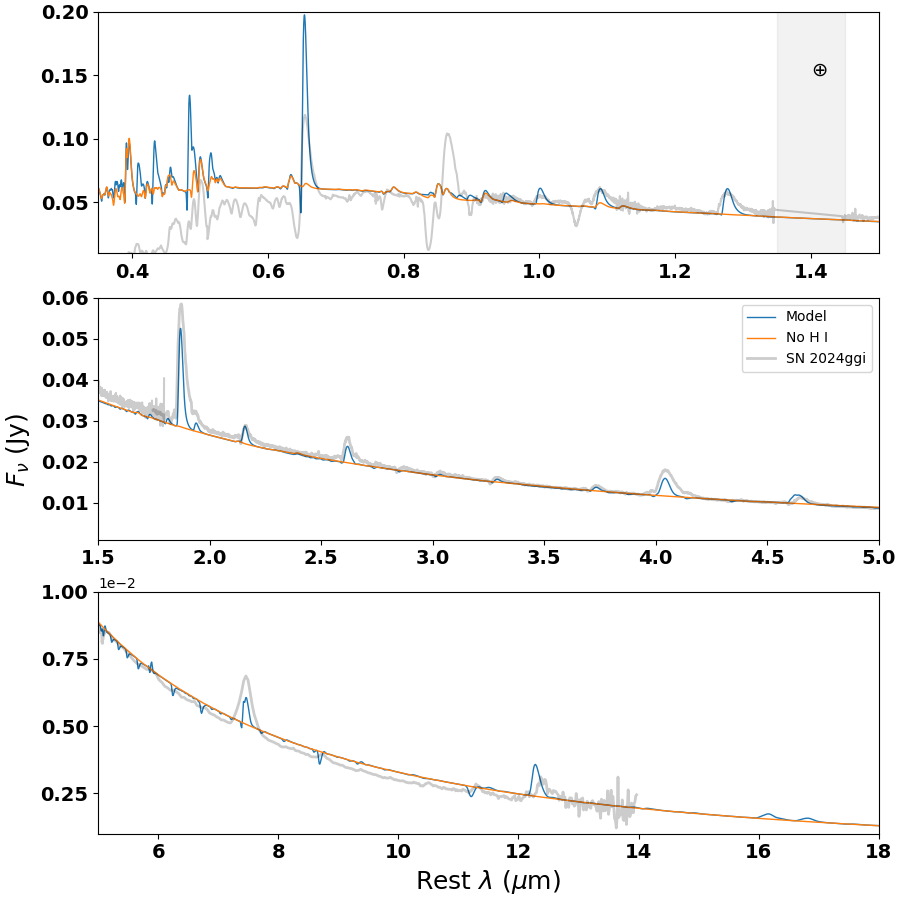}
    \caption{The observed day 55 \ggi \jwst spectrum, 
      the \phx model, and the \phx model with all \ion{H}{1} lines turned
      off. Most of the observed spectral features are due to \ion{H}{1} lines.}
    \label{fig:h1missing}
\end{figure*}

\autoref{fig:h1missing} shows the fiducial model (blue) and the same
model with the
\ion{H}{1} opacities turned off (orange). Beyond the Ca IR triplet
(0.757789, 0.758322, 0.760342~\microns),
there are 
hardly any observed features not accounted for by \ion{H}{1} lines. 

\begin{figure*}[ht]
    \centering
    \includegraphics[width=\textwidth]{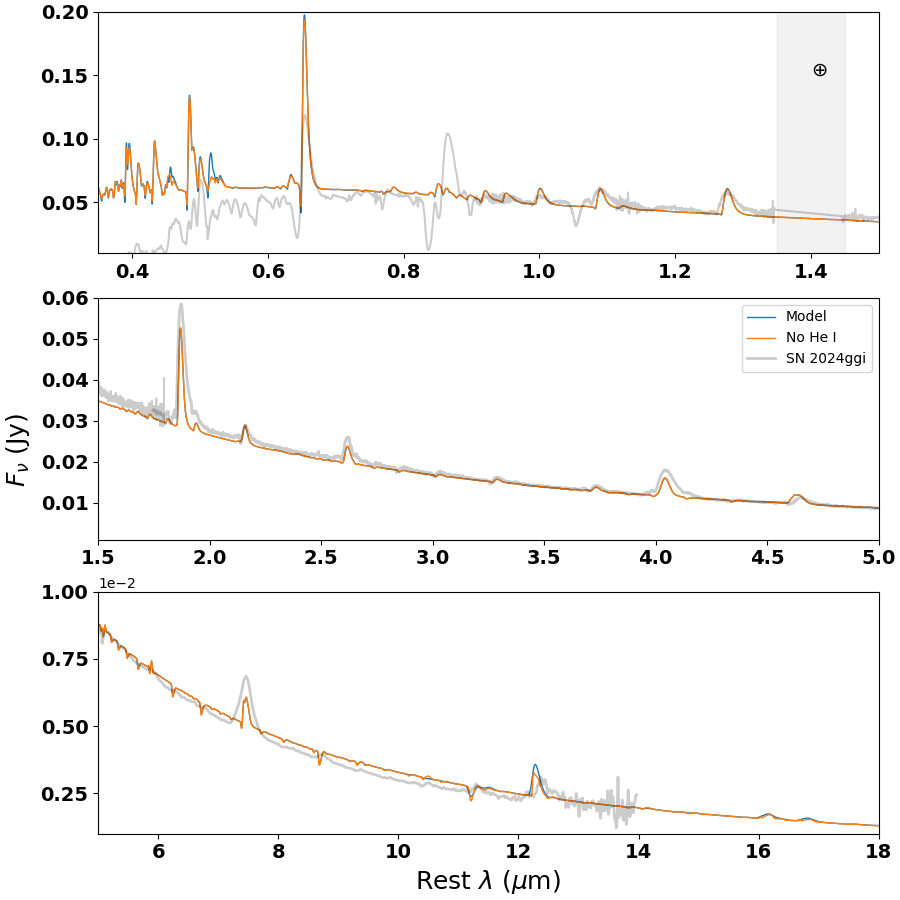}
    \caption{The observed day 55 \ggi ground-based and \jwst spectrum, 
      the \phx model, and the \phx model with all \ion{He}{1} lines turned
      off. The one He I line that is prominent in the model spectrum
      is nearly coincident with the \hl{Hu}{\alpha} line. Thus, there is no
      strong evidence for \ion{He}{1} lines at this epoch and observed
      wavelength range.}
    \label{fig:he1missing}
\end{figure*}

\autoref{fig:he1missing} shows the fiducial model (blue) and the same
model with the
\ion{He}{1} opacities turned off (orange). The only strong features
are the singlet and triplet transitions near 0.50~\microns and the singlet and triplet
transitions near 12.2~\microns.

The only neutral helium lines clearly evident in \autoref{fig:he1missing} are
the 1S-1P $\lambda 5017.1$ line and the  3P-3D multiplet near
12.26~\microns, but the presence of 
\hl{Hu}{\alpha} at 12.372~\microns makes that identification weak. There
are a few other features that contribute somewhat blueward of 0.5~\microns.
It is possible that the feature just blueward of \hl{Pa}{\gamma} is
due to the \ion{He}{1} 1.083\microns line, but the \ion{C}{1}
1.069\microns line seems to be a more likely identification (see \autoref{sec:disc}).
Thus,
the entire observed spectrum is well explained solely by lines
of neutral hydrogen.

\subsection{Line Profiles}
\label{sec:lineprofs}

An obvious inconsistency of the model is that it does not
reproduce the strong observed high velocity component seen in the Ca IR
triplet (and possibly the \ion{Mg}{2} 0.9226 \microns line) in the
optical+NIR (from the 
spectrum obtained on the IRTF), but this is inherent in the model
since the atmosphere contains no material at velocities above
5900~\kmps. 
While the model does an excellent job of fitting the \jwst line profiles
blueward 5~\microns, redward of 5~\microns the model predicts absorption
profiles where the observations show emission profiles.  Although the
\hl{Pf}{\alpha} line is not a great fit in the model it does show strong
net emission as observed. However, for the 7-10 \ion{H}{1} feature at 8.76~\microns,
the observations show net emission, while the model shows a nearly pure
absorption line. This is likely due to a lack of significant optical
depth in the line at the photosphere, but significant optical depth in
the line above the 
photosphere \citep[such that the line is ``detached'', see, for
  example,][]{Branch_etal_2003}. 
In the fiducial  model $\vphot =3000$~\kmps. (In \phx models there is
no pre-defined 
photosphere, so   \vphot is \emph{not}  defined a-priori, as it is in SYNOW.
In \phx we therefore define \vphot as the point where 
$\tau_\text{std} = 1$).
 Examining the departure coefficients
shows the \hl{Pf}{\alpha} levels are both underpopulated, whereas levels 7
and 10 are quite overpopulated to the model surface, making the line
likely to be detached.

\begin{figure}[ht]
    \centering
    \includegraphics[width=0.5\textwidth]{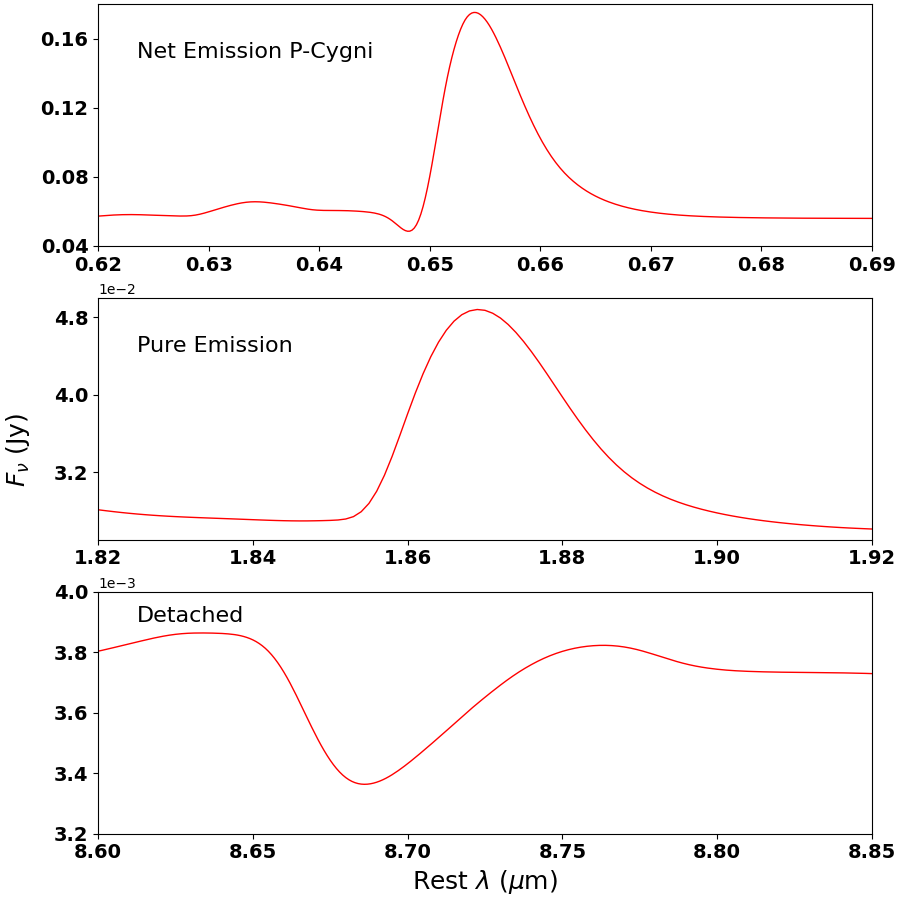}
    \caption{The three different line types seen in the models. Top
      panel: The strongly net-emission P-Cygni \hl{H}{\alpha} line. Middle
      panel: A nearly pure emission line (transition). Bottom panel: a
      P-Cygni line that appears to be pure absorption in
      \autoref{fig:model_data}. The line is essentially detached (see
      \autoref{sec:lineprofs}).}
    \label{fig:lines}
\end{figure}

\autoref{fig:lines} shows the three different line types produced by
the model (and seen in the observations): strongly
net-emission P-Cygni (\hl{H}{\alpha}), a nearly pure emission line, and
a P-Cygni line that appears strongly in absorption. The latter corresponds
to a detached feature. 

\begin{figure*}[ht]
    \centering
    \includegraphics[width=\textwidth]{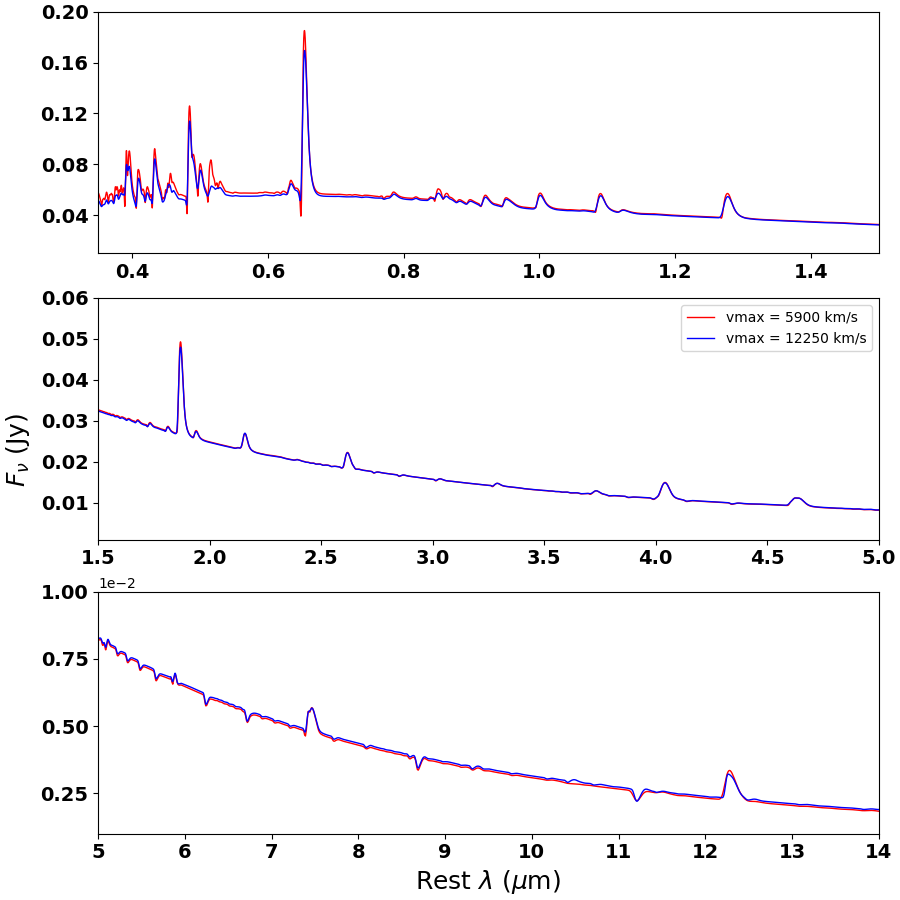}
    \caption{
      A SYNOW fit to the \hl{H}{\alpha} line in \ggi. The optical depth is
      $\tau = 10$, $v_\text{phot} = 3500$~\kmps, the density profile is
      given by an exponential with $v_e = 2000$~\kmps, and
      $v_\text{max} = 20000$~\kmps.
    }
    \label{fig:synow}
\end{figure*}

The \phxO model does a relatively poor job of fitting the blue
absorption troughs of the \ion{H}{1} optical lines, even while fitting
the line widths in the NIR and MIR. In this work we are focused on
using a simple toy model that describes the
NIR + MIR. However, it is worth examining  the velocity extent of
\hl{H}{\alpha}.  \autoref{fig:synow} shows a 
SYNOW calculation of \hl{H}{\alpha}, which does a relatively good job of
fitting the velocity extent of the feature.

The SYNOW parameters are
      $\tau = 10$, $v_\text{phot} = 3500$~\kmps, the density profile is
      given by an exponential with $v_e = 2000$~\kmps, and
      $v_\text{max} = 20000$~\kmps.
This density profile is significantly shallower than the \phxO model,
leading to the more extended blue absorption trough in the P-Cygni
line profile.

\subsection{Velocity Extent}
\label{sec:velex}

\begin{figure*}[ht]
    \centering
    \includegraphics[width=\textwidth]{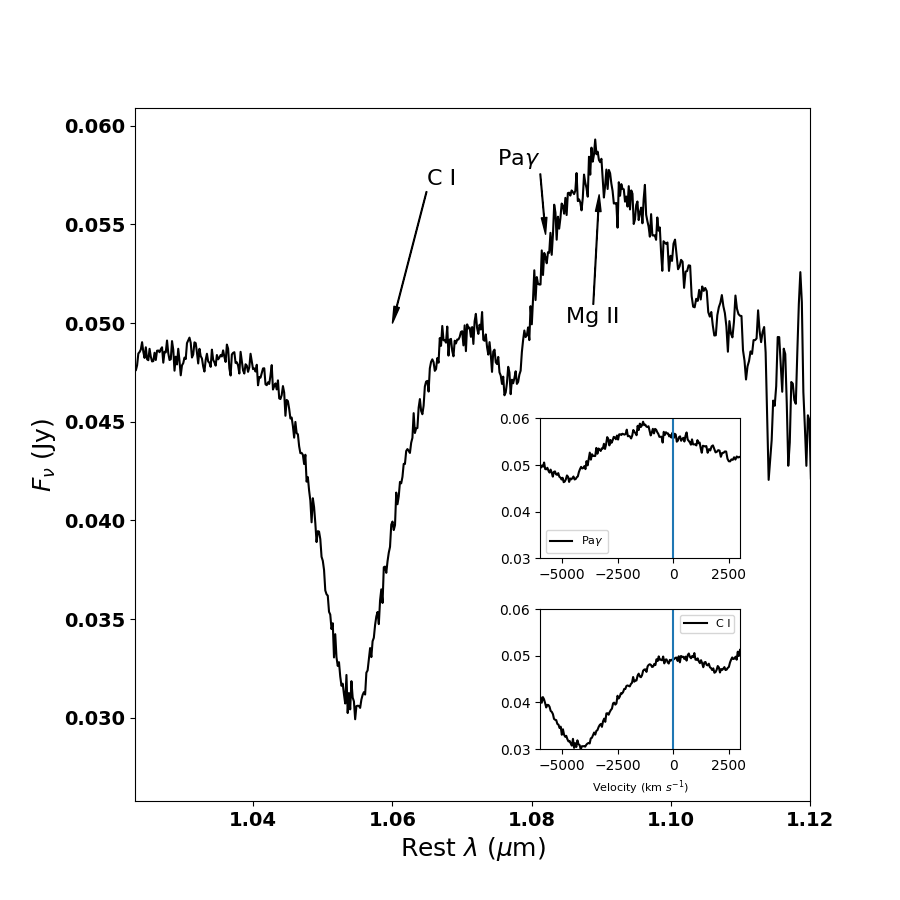}
    \caption{The fiducial model with $\vmax = 5900$~\kmps and
      a model with  $\vmax = 12,250$~\kmps.} 
    \label{fig:vmax}
\end{figure*}

Although there are no indications
of molecular or dust emission at this early phase, it is useful to
know  the velocity extent of the ejecta in order to constrain the
emission radius for future dust. We have run several models in
addition to the fiducial model to explore this. Unfortunately, simply
extending the velocity extent, without altering the density profile
does not affect the predicted spectrum 
significantly enough to draw conclusions about whether the ejecta
could extend to higher velocity. \autoref{fig:vmax} shows the fiducial
model and a model with $\vmax = 12,250$~\kmps.  

\section{Discussion}
\label{sec:disc}

\begin{figure}[ht]
    \centering
    \includegraphics[width=0.5\textwidth]{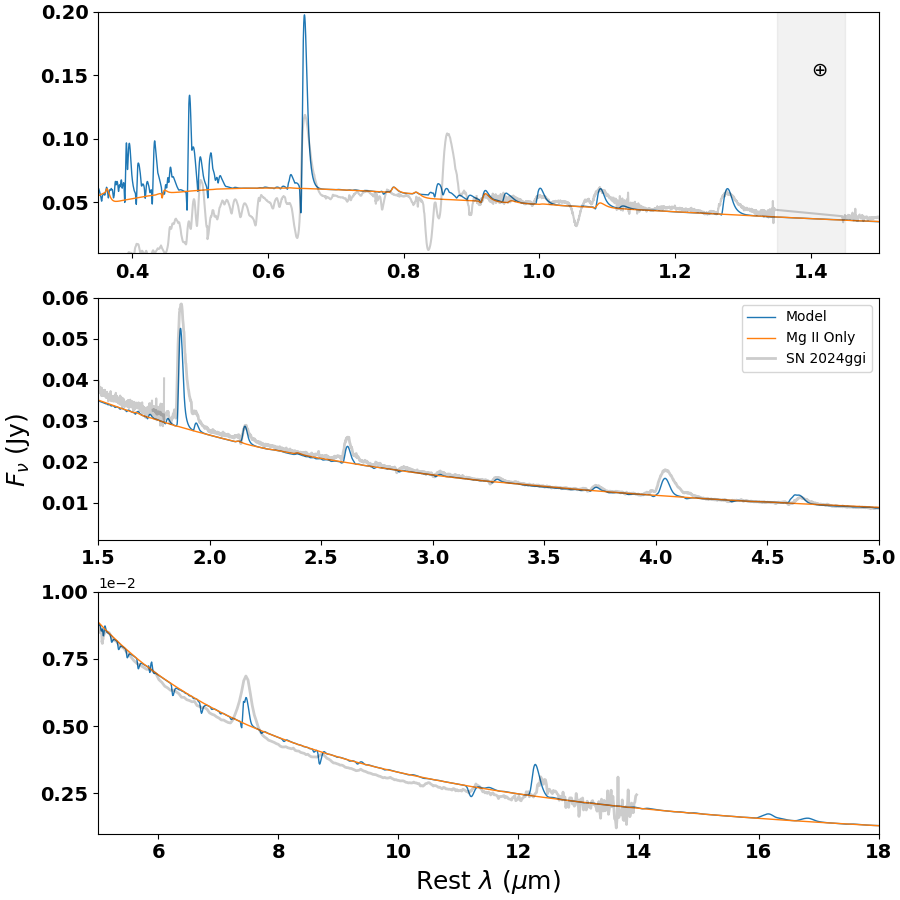}
    \caption{A close up of the P-Cygni features formed by
      \hl{Pa}{\gamma} and the \ion{C}{1}
1.069~\microns line (from the IRTF
      spectrum). Also indicated is the  feature on the  maximum
of \hl{Pa}{\gamma} that is almost certainly due to the \ion{Mg}{2}
1.0926~\microns line (see \autoref{sec:disc}).
The insets show
the velocity extent of the  two P-Cygni lines. 
}
    \label{fig:PagvsCI}
\end{figure}

\begin{figure*}[ht]
    \centering
    \includegraphics[width=\textwidth]{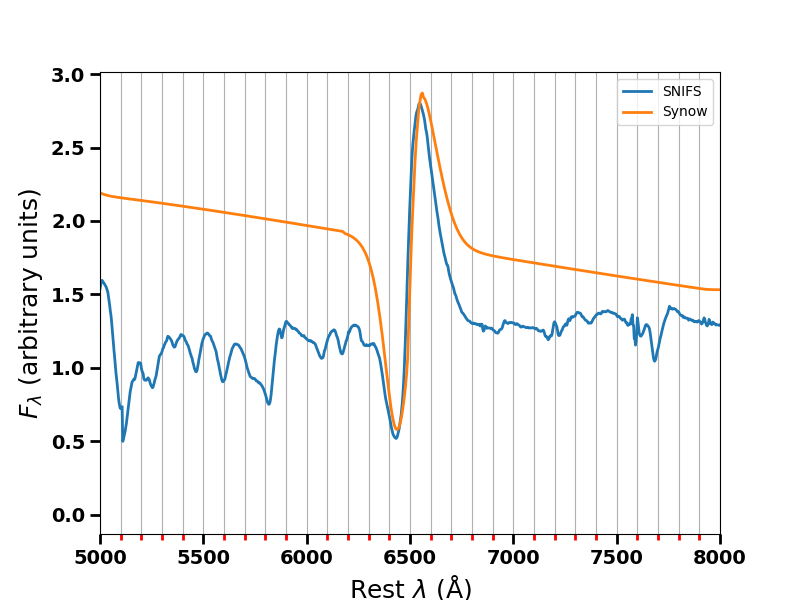}
    \caption{
      Same as \autoref{fig:h1missing} but with all the line
      opacity except for that due to \ion{Mg}{2} turned off.
    }
    \label{fig:mg2_singleion}
\end{figure*}

The observations and modeling show that, with the exception of the Ca
IR triplet and the \ion{C}{1} 1.069~\microns line, all the features in
the NIR+MIR 
spectrum can be attributed to \ion{H}{1}. 
 \autoref{fig:PagvsCI} shows the velocity
extent of the \hl{Pa}{\gamma} line (left) and the \ion{C}{1}
1.069~\microns line (right). They both have roughly the same velocity
($\sim 4,800$~\kmps for \hl{Pa}{\gamma} and  $\sim 4,300$~\kmps for
\ion{C}{1}). The ``bump'' on \hl{Pa}{\gamma} at $v \sim -1300$~\kmps is
interesting, as noted above it is evident in a few other \ion{H}{1} features.
 If the putative
\ion{C}{1} line were instead due to the \ion{He}{1} 1.083\microns
line, it would have a velocity of 7800~\kmps and it would be difficult
to reproduce the P-Cygni shape of the feature since the emission peak
of \ion{He}{1} would be blended with the emission peak of
\hl{Pa}{\gamma}, whereas the \ion{C}{1} line has an emission peak at
its rest wavelength, giving us confidence in the
identification. Nevertheless, the \hl{Pa}{\gamma} feature could be a
blend of \hl{Pa}{\gamma} and \ion{He}{1} that can only be disentangled
by detailed modeling.
 In a study of 30 SNe~II, observed in the NIR,
 \citet{Davis:2019} identify the emission component of \hl{Pa}{\gamma}
 feature at 50~d as  due to \ion{H}{1}, \ion{He}{1}, and the
 \ion{Sr}{2} 1.092~\microns line, with the absorption primarily due
 to \ion{He}{1} 1.083~\microns. Our model, as shown in
 \Autoref{fig:h1missing,fig:he1missing}, finds that the observed 
 feature is due primarily to \hl{Pa}{\gamma} in both the emission and
 absorption components of the P-Cygni feature, and at least in our
 model, there is no strong evidence for the \ion{He}{1} 1.083~\microns
 line. However, \autoref{fig:h1missing} clearly shows that there is an
 additional emission component  
 that is not due to
 either \ion{H}{1} or \ion{He}{1}. We explored the possibility that
 the emission in the model was due to \ion{Sr}{2}, but found no
 evidence to support that. We determined that the model feature is due
 to the \ion{Mg}{2} 1.0926~\microns transition (see
 \autoref{fig:mg2_singleion}).
 Indeed,  \autoref{fig:mg2_singleion} shows significant model contributions
from \ion{Mg}{2} at 0.36376, 0.35521, 0.4432, 0.77899, 0.82316, 0.9226
\microns transitions in addition to the  1.0926~\microns transition. 
Although the model attributes the feature to \ion{Mg}{2}, a small bump
also appears in the observed spectrum at the same velocity in several
other \ion{H}{1} features. This suggests that the observed line
profiles might be partially influenced by another effect. 
These results also show that \ggi would fall into the weak SNe~II
 class of \citet{Davis:2019}. However, \citet{Davis:2019} find that weak
 SNe show an absorption feature that they ascribe to high velocity
 \ion{He}{1}. Since we see no strong evidence
 for the \ion{He}{1} 1.083~\microns feature at this epoch, \ggi does
 not fit well into either of 
 their classes.

The simple, untuned model does a relatively good job of fitting the
strength and width of the lines, suggesting that the most of ejecta is
reasonably well described by a powerlaw density profile, solar
compositions, and a velocity extent $\vmax \la 10,000 $~\kmps.
This result is compatible with the value obtained by \citet{Zhang:2024}.

\subsubsection{Ca-O}

Our model does a poor job on the Ca IR triplet, producing a feature that
is too weak and whose blueshifted absorption is too slow, likely
because the line is just too weak in the model. However, the
model does reproduce the observed separation between the triplet
components. It appears likely that, rather than the IR triplet being
due to  unresolved photospheric and high velocity components $v_{ph} =
2100, v_{HV} = 7200$~\kmps, 
the 8662 and 8498 components produce separate P-Cygni features
with $v = 4900, 4400$~\kmps, respectively. Even though this is not
reproduced by our model \citep[possibly due to
  clumping,][]{Zhang:2024}, the velocity extent is reasonably well
captured by our model. Our model also does not reproduce the
\ion{O}{1} 7773.4~\AA\ line; again the line in the model is much weaker
than in the observed spectrum.  The model also fails to reproduce  the
\ion{Ca}{2} lines that  appear in the two observed spectra as two features that
evolve with time. 

Since the steep power-law density does
not capture the extended velocity profiles in the optical, nor the
line-blanketing, our low estimate of $E_\text{ejecta} \sim 0.5$~foe,
is likely a lower limit to the true value of $E_\text{ejecta}$.

\section{Conclusions}

We present optical to MIR observations of \ggi obtained on the
plateau at $\sim$55~d after first light.
We produce a panchromatic SED from 0.32 to 21~$\micron$. Our data set
includes ground-based optical and NIR spectra as well as both \jwst NIRSpec and MIRI spectra and imaging. \ggi
was the third SN~II to be observed by \jwst in the plateau
phase after SN~2022acko, and SN 2023ixf.  The spectra of \ggi are
consistent with a H recombination spectrum, where the features are
primarily produced by \ion{H}{1} lines.  The NIR+MIR spectral features
are broader and have higher velocities compared to the under-luminous
SN~2022acko. This is consistent with \ggi\ being a  luminous \snii,
likely with a more massive progenitor.

We present a  \phx model which is able to
reproduce many of the observed spectral features. The model has a
density profile with a powerlaw index of $n = -13$. The
velocity extent of the model is rather narrow $\vphot = 3000$~\kmps
 and $\vmax = 5900$~\kmps.  
The main  discrepancy of
the model occurs in the optical, particularly in the line profiles of
\hl{H}{\alpha}, \ion{O}{1} $\lambda 7773.4$, and the Ca IR triplet;
however, the inferred 
velocity of the IR triplet is within that of the fiducial model.
Thus,
it seems likely that clumping is required, perhaps in a cool dense
shell in order to reproduce the Ca IR triplet line profile. 
While a 
SYNOW fit of \hl{H}{\alpha} fits the blueshifted
absorption, simple \phxO models do not. It is beyond
the scope of this work to fit the \hl{H}{\alpha} profile with toy \phxO
models (see \autoref{sec:appB}), but it shows the power of panchromatic studies to validate or
falsify detailed hydrodynamical models.

We see no evidence of pre-existing molecules (CO or SiO) in the system
of \ggi in these early phases. Furthermore, the \phxO model does an
excellent job of reproducing the continuum at 21~\microns
(\autoref{fig:model_data}), showing there is no evidence of warm or cold dust
or excess flux in the MIR. Data we obtained at a later epoch,
  SN 2024ggi
from our JWST Director's Discretionary (DD) programs
DD-6677 and DD-6716 \citep{Ashall_2024_ggi_b,Ashall_2024_ggi_c}
  show clear evidence for CO emission (T.~Evans, in preparation).  Future observations by \jwst
will reveal the formation processes of new molecules and
potentially dust in the system, helping us to elucidate
the origins of dust in the early universe.

Overall, the panchromatic optical to MIR datasets of \sneii provide a powerful tool to investigate the physics of line formation, molecules, and dust in the ejecta. The close proximity of \ggi\ offers a unique opportunity for continued monitoring throughout the lifetime of \jwst. Future comparative studies with SN~2023ixf, which also has a direct progenitor detection and \jwst data, will enable detailed and novel exploration of dust formation pathways and explosion physics in SNe~II in a way not previously possible.

The data is available at MAST: \dataset[doi: 10.17909/50v0-my61]{\doi{10.17909/50v0-my61}}.

\section{Acknowledgments}

E.B., C.A., J.D., M.S., and  P.H. acknowledge support from NASA grants JWST-GO-02114,
JWST-GO-02122, JWST-GO-04522, JWST-GO-04217, JWST-GO-04436,
JWST-GO-03726, JWST-GO-05057, JWST-GO-05290, JWST-GO-06023,
JWST-GO-06677, JWST-GO-06213, JWST-GO-06583. Support for
programs \#2114, \#2122, \#3726, \#4217, \#4436, \#4522,  \#5057,
\#6023, \#6213, \#6583, and \#6677
were provided by NASA through a grant from the Space Telescope Science
Institute, which is operated by the Association of Universities for Research in
Astronomy, Inc., under NASA contract NAS 5-03127.
A.B., E.B., P.B., and P.H. acknowledge support from NASA grant
80NSSC20K0538.
E.B.,  C.A., and J.D. acknowledge support from HST-AR-17555, Support for
Program number 17555 was provided through a grant from the STScI under NASA
contract NAS5-26555.
J.L. acknowledges support from NSF grant AAG-2206523.
L.G. acknowledges financial support from AGAUR, CSIC, MCIN and AEI
10.13039/501100011033 under projects PID2023-151307NB-I00, PIE
20215AT016, CEX2020-001058-M, ILINK23001, COOPB2304, and
2021-SGR-01270. 
M.D. Stritzinger is funded by the Independent Research Fund Denmark (IRFD,
grant number 10.46540/2032-00022B).
Some of this material is based upon work supported by the National Science
Foundation Graduate Research Fellowship Program under Grant
Nos. 1842402 and 2236415. Any opinions, findings, conclusions, or
recommendations expressed in this material are those of the author(s)
and do not necessarily reflect the views of the National Science
Foundation. 
The Shappee group at the University of Hawai'i is supported with funds
from NSF (grants AST-2407205) and NASA (grants HST-GO-17087,
80NSSC24K0521, 80NSSC24K0490, 80NSSC23K1431). 
Some of the calculations presented here were performed at the National Energy
Research Supercomputer Center (NERSC), which is supported by the Office of
Science of the U.S. Department of Energy under Contract
No. DE-AC03-76SF00098.
The authors gratefully acknowledge the computing time
made available to them on the high-performance computers HLRN-IV
at GWDG at the NHR Center {NHR{@}G{\"o}ttingen} and at ZIB at the NHR
Center {NHR{@}Berlin}. These Centers are jointly supported by the Federal
Ministry of Education and Research and the state governments
participating in the NHR (\url{https://www.nhr-verein.de/unsere-partner}). 
We also thank OU
Supercomputing Center for Education \& Research (OSCER) at the University of
Oklahoma (OU). 
The Infrared Telescope Facility is operated by the University of Hawaii under contract 80HQTR24DA010 with the National Aeronautics and Space Administration.

\clearpage

\renewcommand{\thetable}{A\arabic{table}}
\setcounter{table}{0}

\appendix

\section{Log of JWST Spectral Observations}
\label{sec:appA}
\autoref{tab:JWST_spec_info} shows the log of the \jwst spectral observations.

\begin{deluxetable}{cc}
  \tablecaption{Log of the \jwst spectral observations
  \label{tab:JWST_spec_info}} 
  \tablehead{\colhead{Parameter} & \colhead{Value}  }
  \startdata
    \hline
    \multicolumn{2}{c}{NIRSpec Spectral Observations} \\
    \hline
    Mode & Fixed Slit  \\
    $T_{\rm obs}$ (MJD) & 60466.465/60466.460\\
    Phase (days) & 55.52\\     Slit & S400A1 \\
    Subarray & SUBS400A1 \\
    Grating-Filter & G235M-F170LP/G395M-F290LP \\
    Exp Time (s) & 60.82/98.22 \\
    Groups per Integration & 12/20 \\
    Integrations per Exp. & 1/1 \\
    Total Dithers & 3/3 \\
        Total Integrations & 3/3 \\
    Readout Pattern & NRS \\
    \hline
    \multicolumn{2}{c}{MIRI Spectral Observations}  \\
    \hline
    Mode & LRS \\
    $T_{\rm obs}$ (MJD) & 60466.439  \\
    Phase (days) &  55.50 \\      Groups per Integration & 15 \\
    Integrations per Exp. & 3 \\
    Exposures per Dither & 1 \\
    Total Dithers & 2 \\
    Total Exp Time (s) & 260.85 
  \enddata
\end{deluxetable}

\section{\hl{H}{\alpha} line profile}
\label{sec:appB}

\renewcommand{\thetable}{B\arabic{table}}
\setcounter{table}{0}

\renewcommand{\thefigure}{B\arabic{figure}}
\setcounter{figure}{0}

\begin{deluxetable}{lllrrr}
\tablecaption{Parameters of the comparison \phxO models \label{tab:params}}
\tablehead{
\colhead{Model} & \colhead{NLTE} & \colhead{Density Profile}&
\colhead{$v_0$ (\kmps)} & \colhead{$v_e$ (\kmps)} & \colhead{\vmax (\kmps)}
}
\startdata
Fiducial & Yes & $(v/v_0)^{-13}$  &  3000 & \nodata & 5900\\
1 & No &$\exp[-(v-v_0)/v_e]$ & 3500 & 2000 & 13000\\
2 & No &$\exp[-(v-v_0)/v_e]$ & 3500 & 800 & 9800\\
3 & No &$\exp[-(v-v_0)/v_e]$ & 3500 & 800 & 13500
\enddata
\end{deluxetable}

\begin{figure*}[ht]
    \centering
    \includegraphics[width=\textwidth]{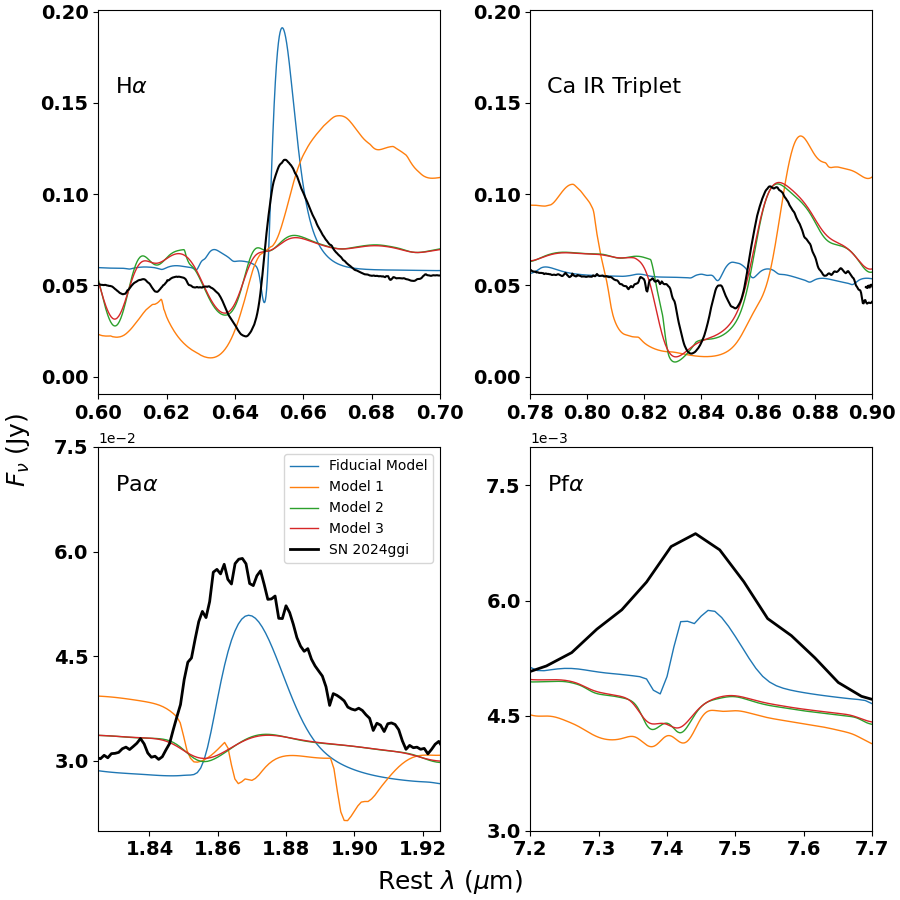}
    \caption{The line profiles of \hl{H}{\alpha} (upper left panel), Ca IR
      triplet (upper right panel) \hl{Pa}{\alpha} (lower left panel), and
      \hl{Pf}{\alpha} (lower right panel). Models with more extended
      velocity profiles capture the velocity extent of the optical
      features, but do a poorer job of reproducing the shape of the
      NIR and MIR features. The synthetic spectra have been scaled for
      clarity. See \autoref{sec:appB} for more details.
    }
    \label{fig:denscomp}
\end{figure*}

In order to  probe the results found with the SYNOW fit to
\hl{H}{\alpha}  a bit further, we ran a few \phxO models in LTE, not
forcing convergence to radiative equilibrium. The point is just
to illustrate the effects of the density profile on the velocity
extent of the lines, not to produce a model that fits every feature
from the optical to the MIR. The model parameters for the test models
(1--3) and the fiducial model are given in
\autoref{tab:params}. \autoref{fig:denscomp} shows the line 
profiles of \hl{H}{\alpha}, the Ca IR triplet, \hl{Pa}{\alpha}, and
\hl{Pf}{\alpha} for \ggi for the four models described in \autoref{tab:params}.
None of the models does a
great job of fitting the blueshifted absorption trough of \hl{H}{\alpha},
Model 1 with the density profile inferred from the SYNOW calculation
extends way too far to the blue, whereas the steeper density profiles
in Models 2 and 3 do a somewhat better job of reproducing the observed
velocity extent. Only the fiducial model does a good job of
reproducing the shape of the rest wavelength \hl{H}{\alpha} emission
peak. Moving to the Ca IR triplet, all of the LTE models produce
blueshifted absorption troughs that extend too far to the blue and show no
separation among the components of the triplet, whereas that is
evident in the data. The steeper density Models 2 and 3 do a
reasonable job of replicating the rest wavelength emission peak of
the P-Cygni profile, while the fiducial model shows only a very weak
Ca IR triplet. Moving to the NIR and MIR the fiducial model does a
much better job of reproducing both \hl{Pa}{\alpha} and \hl{Pf}{\alpha}.

While it is beyond the scope if this work to attempt to find ``the
model'' of \ggi, it is clear that the fiducial model captures much of
the line and continuum formation in the NIR and MIR, but that there is
more complexity needed to reproduce the optical features. This work
does show the power of long multi-wavelength spectroscopy and
photometry to produce data that can validate or falsify detailed models.

\clearpage

\bibliography{sn24ggi_bib}

\end{document}